\documentclass[final,times,5p]{elsarticle}
\usepackage{graphicx}
\usepackage{xspace}
\usepackage{amsmath}
\usepackage{amsfonts}
\usepackage{psfrag}
\usepackage[psamsfonts]{amssymb}
\usepackage[dvips]{color}
\usepackage{srcltx}
\usepackage{hyperref}
\DeclareMathOperator{\Real}{Re}
\DeclareMathOperator{\Imag}{Im}

\newcommand{\JETSET}{JETSET}
\newcommand{\SKIP}[1]{}
\newcommand{\ee}{\ensuremath{e^{+}e^{-}}\xspace}

\newcommand{\PP}{\ensuremath{\psi(2S)}\xspace}
\newcommand{\JP}{\ensuremath{J/\psi}\xspace}
\newcommand{\DP}{\ensuremath{\psi(3770)}\xspace}

\newcommand{\MPP}{M\xspace}

\newcommand{\GPP}{\ensuremath{\Gamma}\xspace}

\renewcommand{\Re}{\ensuremath{\text{Re}}\,}
\renewcommand{\Im}{\ensuremath{\text{Im}}\,}
\renewcommand{\epsilon}{\varepsilon}

\newcommand{\AP}{$\approx$}
\newcommand{\BL}{\phantom{$\approx$}}

\newcommand{\LT}{$<$}

\newcommand{\GGG}{\ensuremath{\Gamma_{ee}\times\mathcal{B}_h\xspace}}

\newcommand{\Gh}{\ensuremath{\tilde{\Gamma}_h}\xspace}
\newcommand{\B}{\mathcal{B}}

\newcommand{\TNL}{\\*[1pt]}

\begin{document}
\begin{frontmatter}
\title{Measurement of main parameters of the \PP resonance}
\author[binp]{V.\,V.\,Anashin}
\author[binp,nsu]{V.\,M.\,Aulchenko}
\author[binp,nsu]{E.\,M.\,Baldin} 
\author[binp]{A.\,K.\,Barladyan}
\author[binp]{A.\,Yu.\,Barnyakov}
\author[binp]{M.\,Yu.\,Barnyakov}
\author[binp,nsu]{S.\,E.\,Baru}
\author[binp]{I.\,Yu.\,Basok}
\author[binp,nsu]{O.\,L.\,Beloborodova}
\author[binp]{A.\,E.\,Blinov}
\author[binp,nstu]{V.\,E.\,Blinov}
\author[binp]{A.\,V.\,Bobrov }
\author[binp]{V.\,S.\,Bobrovnikov}
\author[binp,nsu]{A.\,V.\,Bogomyagkov}
\author[binp,nsu]{A.\,E.\,Bondar}
\author[binp]{A.\,R.\,Buzykaev}
\author[binp,nsu]{S.\,I.\,Eidelman}
\author[binp]{D.\,N.\,Grigoriev}
\author[binp]{Yu.\,M.\,Glukhovchenko}
\author[binp]{V.\,V.\,Gulevich}
\author[binp]{D.\,V.\,Gusev}
\author[binp]{S.\,E.\,Karnaev}
\author[binp]{G.\,V.\,Karpov}
\author[binp]{S.\,V.\,Karpov}
\author[binp,nsu]{T.\,A.\,Kharlamova}
\author[binp]{V.\,A.\,Kiselev}
\author[binp]{V.\,V.\,Kolmogorov}
\author[binp,nsu]{S.\,A.\,Kononov}
\author[binp]{K.\,Yu.\,Kotov}
\author[binp,nsu]{E.\,A.\,Kravchenko}
\author[binp]{V.\,N.\,Kudryavtsev}
\author[binp,nsu]{V.\,F.\,Kulikov}
\author[binp,nstu]{G.\,Ya.\,Kurkin}
\author[binp,nsu]{E.\,A.\,Kuper}
\author[binp,nstu]{E.\,B.\,Levichev}
\author[binp,nsu]{D.\,A.\,Maksimov}
\author[binp]{V.\,M.\,Malyshev}
\author[binp]{A.\,L.\,Maslennikov}
\author[binp,nsu]{A.\,S.\,Medvedko}
\author[binp,nsu]{O.\,I.\,Meshkov}
\author[binp]{S.\,I.\,Mishnev}
\author[binp]{I.\,I.\,Morozov}
\author[binp]{N.\,Yu.\,Muchnoi}
\author[binp]{V.\,V.\,Neufeld}
\author[binp]{S.\,A.\,Nikitin}
\author[binp,nsu]{I.\,B.\,Nikolaev}
\author[binp]{I.\,N.\,Okunev}
\author[binp,nstu]{A.\,P.\,Onuchin}
\author[binp]{S.\,B.\,Oreshkin}
\author[binp,nsu]{I.\,O.\,Orlov}
\author[binp]{A.\,A.\,Osipov}
\author[binp]{S.\,V.\,Peleganchuk}
\author[binp,nstu]{S.\,G.\,Pivovarov}
\author[binp]{P.\,A.\,Piminov}
\author[binp]{V.\,V.\,Petrov}
\author[binp]{A.\,O.\,Poluektov}
\author[binp]{V.\,G.\,Prisekin}
\author[binp]{A.\,A.\,Ruban}
\author[binp]{V.\,K.\,Sandyrev}
\author[binp]{G.\,A.\,Savinov}
\author[binp]{A.\,G.\,Shamov\corref{cor}}
\author[binp]{D.\,N.\,Shatilov}
\author[binp,nsu]{B.\,A.\,Shwartz}
\author[binp]{E.\,A.\,Simonov}
\author[binp]{S.\,V.\,Sinyatkin}
\author[binp]{A.\,N.\,Skrinsky}
\author[binp]{V.\,V.\,Smaluk}
\author[binp]{A.\,V.\,Sokolov}
\author[binp]{A.\,M.\,Sukharev}
\author[binp,nsu]{E.\,V.\,Starostina}
\author[binp,nsu]{A.\,A.\,Talyshev}
\author[binp,nsu]{V.\,A.\,Tayursky}
\author[binp,nsu]{V.\,I.\,Telnov}
\author[binp,nsu]{Yu.\,A.\,Tikhonov}
\author[binp,nsu]{K.\,Yu.\,~Todyshev\corref{cor}}
\cortext[cor]{Corresponding authors, e-mails: \\ shamov@inp.nsk.su,~~todyshev@inp.nsk.su }
\author[binp]{G.\,M.\,Tumaikin}
\author[binp]{Yu.\,V.\,Usov}
\author[binp]{A.\,I.\,Vorobiov}
\author[binp]{A.\,N.\,Yushkov}
\author[binp]{V.\,N.\,Zhilich}
\author[binp,nsu]{V.\,V.\,Zhulanov}
\author[binp,nsu]{A.\,N.\,Zhuravlev} 

 \address[binp]{Budker Institute of Nuclear Physics, Russian Acad. Sci. 
Sibirean Div.,630090, Novosibirsk, Russia}
  \address[nsu]{Novosibirsk State University, 630090, Novosibirsk, Russia}
  \address[nstu]{Novosibirsk State Technical University, 630092,
    Novosibirsk, Russia}

  \begin{abstract}
A high-precision determination of the main parameters of the \PP
resonance has been performed 
with the KEDR detector at the VEPP-4M \ee collider in three scans 
of the \mbox{$\psi(2S)$--\DP} energy range.
Fitting the energy dependence of the multihadron cross section
in the vicinity of the \PP we obtained \textcolor{black}{the mass value
\begin{equation*}
  M = 3686.114 \pm 0.007 \pm 0.011 \:\,^{+0.002}_{-0.012} \:\:\text{MeV}
\end{equation*}}
and the product of the electron partial width by the branching fraction 
into hadrons
\begin{equation*}
  \begin{split}
    \GGG&=      2.233 \pm 0.015  \pm 0.037 \pm 0.020 \, \text{keV}\,. \\
  \end{split}
\end{equation*}
\textcolor{black}{The third error quoted is an estimate of the
\textcolor{black}{model dependence of the result}
due to assumptions on the interference effects
in the cross section of \textcolor{black}{the single-photon} $e^{+}e^{-}$
annihilation to hadrons explicitly
considered in this work. 
Implicitly, the
same assumptions were employed to obtain the charmonium
leptonic width and the absolute branching fractions
in many experiments.}

Using \textcolor{black}{the result presented and}
the world average values of the electron and hadron
branching fractions, one obtains the electron partial width  and the
total width of the \PP:
\begin{equation*}  
 \begin{split}
\Gamma_{ee}&=2.282 \pm 0.015 \pm 0.038 \pm 0.021 \,\text{keV}, \\
\Gamma & = 296  \pm 2 \pm 8 \pm 3 \, \text{keV}.
 \end{split}
\end{equation*}
These results are consistent with and 
more than two times more precise than any  of the previous experiments.
\end{abstract}

\end{frontmatter}
\section{Introduction}
\label{sec:intro}
More than thirty six years passed since the discovery of \JP, but 
studies of charmonium states still raise new questions. Recent 
progress in charmonium physics requires an improvement of the accuracy
of the parameters of charmonium states~\cite{nora}.
This \textcolor{black}{Letter} describes a measurement of the \PP meson
parameters in the KEDR experiment performed during energy scans 
from 3.67 to 3.92 GeV at the VEPP-4M \ee collider. 

For a precision experiment it is essential to state explicitly
what quantities are measured and how they can be compared
with results of theoretical studies, therefore we discuss
a definition of the \PP parameters just after a brief description
of the experiment. The importance of the question has
grown since the appearance of 
\textcolor{black}{the work~\cite{BESelwid} in which
the BES collaboration used an original approach
to the determination of the $J^{PC}\!=1^{- -}$ resonance parameters.
Its further modification has been used in 
Refs.~\cite{Ablikim:2006zq,Ablikim:2007gd,Ablikim:2008zzb}.}

\section{VEPP-4M collider and KEDR detector}
\label{sec:VEPP}
VEPP-4M is an \ee collider~\cite{Anashin:1998sj} designed for high-energy
physics experiments in the center-of-mass (c.m.) energy range from 2 to
12 GeV. The peak luminosity in the 
$2\!\times\!2$ bunches
 operation mode is
about~\(2\!\times\!10^{30}\,\text{cm}^{-2}\text{s}^{-1}\)
in the vicinity of \PP. Having a modest luminosity,
VEPP-4M is well equipped for  high-precision measurements of 
beam energy~\cite{Blinov:2009}.
The instantaneous value of the beam energy can be calibrated
using the resonant depolarization method 
(RDM)~\cite{Bukin:1975db,Skrinsky:1989ie} with
the relative accuracy of about \(10^{-6}\). 
The results of RDM calibrations must be interpolated to determine
the energy during data taking and the interpolation accuracy 
of about 10~keV can be reached~\cite{Aulchenko:2003qq}.
Continuous energy monitoring is performed using 
the infrared light Compton backscattering~\cite{CBS} with 
the accuracy of the method $\sim$ 60~keV.

\begin{figure}[t]
\includegraphics[width=0.49\textwidth]{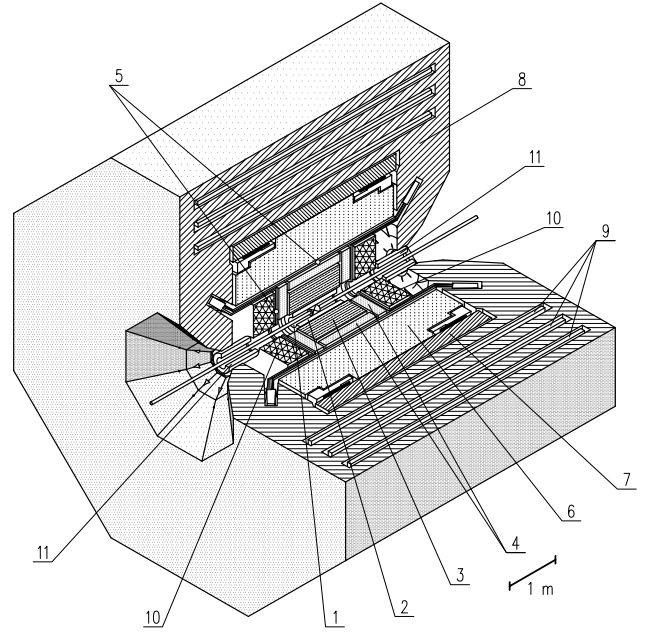}
\caption{{\normalsize 1~--~Vacuum chamber, 2~--~Vertex detector, 
3~--~Drift chamber,
4~--~Threshold aerogel counters, 5~--~ToF-counters, 
6~--~ Liquid krypton calorimeter, 7~--~Superconducting solenoid, 
8~--~Magnet yoke,9~--~Muon tubes, 10~--~CsI calorimeter,
11~--~Compensating superconducting coils}}

\label{kedr_picture}
\end{figure}

The KEDR detector~\cite{Anashin:2002uj} (Fig.~\ref{kedr_picture})
comprises the vertex detector (VD), 
drift chamber (DC),  time-of-flight (TOF) system of scintillation counters,
particle identification system based on the aerogel Cherenkov
counters, EM calorimeter (liquid krypton in the barrel part and
CsI crystals in the endcaps), superconducting magnet system 
and muon system inside the magnet yoke. The superconducting 
solenoid provides a longitudinal magnetic field of 0.6~T.
The detector is equipped with a scattered electron tagging system
for two-photon studies and some applications.
The on-line luminosity measurement is provided by  two independent
single bremsstrahlung monitors.
The trigger consists of two hardware levels:
the primary trigger (PT) and the secondary one (ST)~\cite{TALYSHEV}.
The PT operates using signals from the TOF counters and fast signals
from the CsI and LKr calorimeters,
whereas the ST uses the normally shaped calorimeter signals
and the information from VD, DC and TOF system.
After the readout, a software selection of events
is performed using simplest event characteristics, in particular,
the number of hits in VD. The upper limit on the number of VD
tubes hitted is very effective for the machine background suppression.
\section{Data sample}
\label{sec:Exp}
In 2004 two scans of the {\PP}--{\DP} energy range  were carried out
with an integrated luminosity of about 0.7~$\text{pb}^{-1}$. 
In 2006 the regions of \PP and \DP mesons were scanned once again
with an integrated luminosity of  $\approx$ 1.9~pb$^{-1}$.  
The combined data sample corresponds to $1.6 \cdot 10^5$  \PP produced.
The data acquisition scenario for \PP was similar to that 
described in Ref.~\cite{Aulchenko:2003qq}.
The accuracy of the energy interpolation between the RDM calibrations
varied from 10 to 30 keV during the whole experiment.
\textcolor{black}{Our final results
on the \DP parameters are presented in the next letter of this volume.}

\section{On definition of $J^{PC}\!=1^{- -}$ resonance parameters}
\label{sec:param}

A resonance with the quantum numbers $J^{PC}\!=1^{- -}$ can be treated
in some cases not as an unstable particle but just as a vacuum polarization
phenomenon. Sometimes this causes confusion in the data analysis as was
noted in Ref.~\cite{PHIPSI}. To avoid confusion,  ``bare'' and
``dressed'' or ``physical'' parameters of a resonance must be clearly
distinguished.
The former do not include QED corrections and are used in many
theoretical studies, the latter include some of them 
(in particular, the vacuum polarization)
and are published as results of almost all
experimental papers.

The physical parameters correspond
to the interpretation of a $1^{- -}$ resonance as a particle
described with a Breit-Wigner amplitude representing its
appearance in all orders
of the QED perturbations \textcolor{black}{, which is
absolutely natural for the strong production of a resonance}.
For an electromagnetic production of a resonance
like, e.g.,  in the process
$\ee\!\to\PP\!\to\!p\overline{p}$,
this amplitude interferes with the
pure QED amplitude of the $\ee\!\to\!p\overline{p}$ transition. 
\textcolor{black}{As shown below, such an approach 
allows one to avoid the numerical integration in the calculation of
radiative corrections.}
\SKIP{For an electromagnetic production of a resonance
like, e.g.,  in the process
$\ee\!\to\PP\!\to\!p\overline{p}$,
this amplitude interferes with the
pure QED amplitude of the $\ee\!\to\!p\overline{p}$ transition. A
Breit-Wigner amplitude with the physical mass and width can be also applied
to hadronic production like in the process  
$p\overline{p}\!\to\PP\!\to\!p\overline{p}$,
when there are too few reasons 
to consider the vacuum polarization explicitly,
however, that must be done if the bare parameters are employed in the analysis.}

Let us demonstrate the relation between the bare and physical
parameters.
According to Ref.~\cite{KF}, the cross section of the
single--photon annihilation
can be written in the form
\begin{equation}\label{eq:RadCorInt}
  \sigma(s) = \int\!dx\,
     \frac{\sigma_0((1\!-\!x)s)}{|1-\Pi((1\!-\!x)s)|^2}\, \mathcal{F}(s,x),
\end{equation}
where $s$ is the c.m. energy squared, $\mathcal{F}(s,x)$ is the
radiative correction function,
$\Pi(s)$ represents the vacuum polarization operator
and $\sigma_0(s)$ is the Born cross section of
the process.
One has $\Pi = \Pi_0 + \Pi_{R}$ with the nonresonant
$\Pi_0=\Pi_{ee}\!+\!\Pi_{\mu\mu}\!+\!\Pi_{\tau\tau}\!+\!\Pi_{\text{hadr}}\,$
and the resonant
\begin{equation}
\Pi_R(s) = \frac{3\Gamma^{(0)}_{ee}}{\alpha}\frac{s}{M_0}
       \frac{1}{s-M_0^2+i M_0\Gamma_0}\,, 
\label{eq:PiR}
\end{equation}
where \textcolor{black}{$\alpha$ is the fine structure constant}, 
$M_0$, $\Gamma_0$ and $\Gamma^{(0)}_{ee}$ are the ``bare'' resonance mass, 
total and electron widths, respectively. 
\textcolor{black}{Eq.~\eqref{eq:PiR} slightly differs from the expression used
in Refs.~\cite{Berends:1976zp,Burkhardt:1989ky}. It  corresponds to the
simplest resonance cross section parameterization
\begin{equation}
 \sigma_{R}(s) = -\frac{4\alpha}{s}\,\Im\Pi_R(s)  =
 \frac{12\pi\Gamma^{(0)}_{ee}\Gamma_0}{(s-M_0^2)^2+M_0^2\Gamma_0^2}\,. 
\end{equation}}

For the muon pair production the Born cross section
at $s\gg 4m_{\mu}^2\,$ ($m_{\mu}$ is the muon mass)  is just
\begin{equation}
  \sigma^{\mu\mu}_0(s) = \frac{4\pi\alpha^2}{3s},
\label{eq:BornMM}
\end{equation}
thus the resonance behaviour of the cross section is, in this approach,
entirely due to the vacuum polarization, which implicitly describes
the muonic decay of a resonance.
Eqs.~\eqref{eq:RadCorInt}, \eqref{eq:PiR} and
\eqref{eq:BornMM} give the
dimuon cross section without separation into
the continuum, resonant and interference parts.
To obtain the contribution of the resonance,
the continuum part must be subtracted from the amplitude. It can be done
with the identity
\begin{equation}\label{eq:Ident}
\begin{split}
  & \frac{1}{1\!-\!\Pi_0\!-\!\Pi_R(s)} \equiv \\
  & \:\:\:\:\:\:\:\:\:\:\:\:\:\:\:\:\:\:\:\:
  \frac{1}{1\!-\!\Pi_0} +
  \frac{1}{(1\!-\!\Pi_0)^2}\,
  \frac{3\Gamma^{(0)}_{ee}}{\alpha}\,\frac{s}{M_0}\,
  \frac{1}{s\!-\!\tilde{M}^2+i \tilde{M}\tilde{\Gamma}}\,
\end{split}
\end{equation}
Two terms in the right-hand side correspond
to the continuum amplitude and the resonant one, respectively.
\textcolor{black}{The second power of
the vacuum polarization factor $1/(1-\Pi_0)$ in the latter
can be interpreted as the  presence of two photons, one at a resonance
production and the other in its decay.
In the resonant amplitude}
both $\tilde{M}$ and $\tilde{\Gamma}$ depend on $s$
\begin{equation}
\begin{split}
& \tilde{M}^2 = M_0^2 + \frac{3\Gamma^{(0)}_{ee}}{\alpha}\,\frac{s}{M_0}\:
                \Re\frac{1}{1-\Pi_0} \,, \\
& \tilde{M}\tilde{\Gamma} = M_0\Gamma_0 -
  \frac{3\Gamma^{(0)}_{ee}}{\alpha}\,\frac{s}{M_0}\: \Im\frac{1}{1-\Pi_0} \,.
\end{split}
\label{eq:Dress}
\end{equation}
In the vicinity of a narrow resonance this dependence is negligible, thus
the resonant contribution can be described with a simple
Breit-Wigner amplitude
containing the physical parameters $M\approx\tilde{M}(M_0^2)$ and
$\Gamma\approx\tilde{\Gamma}(M_0^2)$.

\textcolor{black}{To obtain the dimuon cross section one has to
multiply the absolute value squared of the right-hand side 
of Eq.~\eqref{eq:Ident}
by the Born cross section~\eqref{eq:BornMM}.
The resonant part of the cross section is proportional to
$\Gamma^{(0)}_{ee}$ squared which appears
instead of $\Gamma^{(0)}_{ee}\Gamma^{(0)}_{\mu\mu}$
due to the lepton universality in QED.}
The factor $1/(1-\Pi_0)^2$ in front of the resonance amplitude
converts the square of the bare $\Gamma^{(0)}_{ee}$
to the square of the physical partial width
\begin{equation}
\Gamma_{ee}=\frac{\Gamma^{(0)}_{ee}}{|1-\Pi_0|^2}
\label{eq:GeePhys}
\end{equation}
\textcolor{black}{recommended to use by Particle Data Group since
the work~\cite{Alexander:1988em} appeared}.

For electromagnetic decays of a resonance to hadrons,
the only difference with the dimuon
case is the factor $R$ in the cross section, where $R$ is the hadron-to-muon
cross section ratio off the resonance peak. 
\textcolor{black}{For a strong decay with the partial
width $\Gamma^{(s)}_0$ the Born cross section is}
\begin{equation}
  \sigma^{(s)}_0(s) =
 \frac{12\pi\,\Gamma^{(0)}_{ee}\,
        \Gamma_0^{(s)}}{(s-M_0^2)^2+M_0^2\Gamma_0^2}\,.
\label{eq:Born3G}
\end{equation}
In this case the identity \eqref{eq:Ident} is not required, the
direct substitution of \eqref{eq:PiR} and \eqref{eq:Born3G} in
\eqref{eq:RadCorInt} leads
to the same definition of the physical mass, total width and leptonic width.
\textcolor{black}{The equivalent definition of the physical mass
in the hadronic channel is given in Ref.~\cite{DZhang}.}
The physical value of the partial width $\Gamma^{(s)}$ is identical to the
bare one $\Gamma_0^{(s)}$.

We would like to emphasize
that the experimental values of a $1^{- -}$ resonance
mass, total  or leptonic width can not be compared with
the immediate results of potential models or used to fit
parameters of a potential without either
``undressing'' of the experimental values or ``dressing'' of the
potential model results with Eqs.~\eqref{eq:Dress} and \eqref{eq:GeePhys}.
The differences between dressed and bare masses are about 1.2 and 0.5~MeV
for the $J/\psi$ and  \PP, respectively. \textcolor{black}{The corresponding
differences for the total widths are about 23 and 10~keV
($\Im\Pi_0\!\approx\!-\alpha\,(R\!+\!2)/3\,$ with the $R$ ratio about 2.2).}

Unlike the \textcolor{black}{works~\cite{BESelwid,Ablikim:2006zq}},
we consistently use the physical parameters and treat equally 
strong and electromagnetic decays of the \PP. The difference in the
approaches is discussed in more detail in Section~\ref{sec:OMHXS}.
\textcolor{black}{It is worth noting that we do not suggest any
new approach, but just follow the one employed in most measurements
of heavy quarkonium parameters though, as far as we know, its relation 
to Eq.~\eqref{eq:RadCorInt} in the hadronic and leptonic channels
was not rigorously considered until recently.}

\section{Cross section calculation}
\label{sec:theory}

The cross section formulae given below in this section
contain the \PP total width $\Gamma$ and the electronic width
$\Gamma_{ee}$, the determination of which are the goal of our analysis.
We fix these parameters in the cross section fit, but use
the iteration procedure to obtain the final results.
The values of the \PP  electron and hadron branching fractions, 
required to recalculate \GGG\, to $\Gamma_{ee}$ and $\Gamma$, are
fixed at the world averages.
The systematic uncertainties due to such an approach
are discussed in Section~\ref{err:fit}.

\subsection{Multihadron cross section}

\textcolor{black}{Below we present the
results of the paper~\cite{Azimov:eq} published soon after
$J/\psi$ discovery in the updated interpretation. Some details
of the analytical calculations and numerical checks
can be found in Ref.~\cite{Todyshev:eq}.}

Using the physical values of the parameters, for  strong
decays of \PP one reduces Eq.~\eqref{eq:RadCorInt} to
\begin{equation}
\begin{split} 
\sigma^{RC}_{\PP}(W)& = \int \frac{12\pi\,\Gamma_{ee}\Gamma^{(s)}_h} 
{\left(W^2(1\!-\!x)-\MPP^2\right)^2+\MPP^2 \Gamma^2} \\
\times&~\mathcal{F}(x,W^2)~dx \, ,
\end{split} 
\label{PPRes:eq}
\end{equation}
where $W\!=\!\sqrt{s}$ is the total collision energy, 
$\Gamma,~\Gamma_{ee}$ and $\Gamma^{(s)}_{h}$ are the total and partial
widths of the \PP meson, and $\,\MPP$ is its mass.

Taking into account the resonance-continuum interference and performing
the integration over $x$ with a simplified version of $\mathcal{F}(x,s)$
one obtains
\begin{equation}
\begin{split} 
\sigma^{RC}_{\PP}(W) &= \frac{12\pi} { W^2 }
 \Bigg\{ \bigg(1+\delta_{\mathrm{sf}}\bigg)
\Bigg[
  \frac{\Gamma_{ee}\Gh}{\GPP \MPP} 
\Imag{f(W)} \\
& -\,\frac{2 \alpha\sqrt{R\,\Gamma_{ee}\Gh\,}}{3 W}\,
\lambda\,\Real{\frac{f^{*}(W)}{1\!-\!\Pi_0}}\
\Bigg]\\
&-\,\frac{\beta\,\Gamma_{ee}\Gh}{2 \Gamma \MPP}\,
\Bigg[
 \Bigg(1\!+\!\frac{\MPP^2}{W^2}\Bigg)\,
\arctan{\frac{\Gamma W^2}{M(M^2\!-\!W^2\!+\!\Gamma^2)}} \\
  &-\frac{\GPP \MPP}{2 W^2}
  \ln{\frac{\bigg(\frac{\MPP^2}{W^2}\bigg)^2+\bigg(\frac{\GPP
        \MPP}{W^2}\bigg)^2}{\bigg(1-\frac{\MPP^2}{W^2}\bigg)^2+\bigg(\frac{\GPP
        \MPP}{W^2}\bigg)^2}}
 \Bigg]\,
 \Bigg\}\,,
\end{split} 
\label{BWrelativistic}
\end{equation}
where 
$\Pi_0$ is the vacuum polarization operator with the \PP contribution
excluded. \textcolor{black}{The $\Gh$ parameter includes
both strong and electromagnetic decays and some
contribution of interference effects which is discussed in the
next subsection}.

The first square bracket in Eq.~\eqref{BWrelativistic} corresponds to
radiation of soft photons, while  the second one represents hard photon 
corrections. 
The $\lambda$ parameter introduced in
Ref.~\cite{Azimov:eq} characterizes the strength of the interference
effect in the multihadron cross section and equals 1 for the dimuon 
cross section.

The correction $\delta_{\text{sf}}$ follows from the 
structure function approach of Ref.~\cite{KF}: 
\begin{equation}\label{eq:deltasf}
  \delta_{\mathrm{sf}}=\frac{3}{4}\beta+
   \frac{\alpha}{\pi}\left(\frac{\pi^2}{3}-\frac{1}{2}\right)+
  \beta^2\left(\frac{37}{96}-\frac{\pi^2}{12}-
  \frac{1}{36}\ln\frac{W}{m_e} \right)\,,
\end{equation}
\begin{equation}
\beta = \frac{4\alpha}{\pi} \left( \ln\frac{W}{m_e} -\frac{1}{2}\right)\,,
\label{eq:beta}
\end{equation}
$m_e$ is the electron mass and the function $f$ is defined 
\textcolor{black}{with}
\begin{equation}
f(W) =  \frac{\pi\beta}{\sin{\pi\beta}}\,
   \Bigg(\frac{W^2}{\MPP^2-W^2-i \MPP \GPP} \Bigg)^{1-\beta}
\!\!\!\!\!.
\label{resfunction}
\end{equation}
The presentation of the soft photon integrals
in form of the real and imaginary
parts of the function $f$ \textcolor{black}{is} more transparent
than that of Ref.~\cite{Cahn:1986qf}.

Despite a simplification of $\mathcal{F}(x,s)$, not too far from 
the \PP peak the resonant part
of Eq.~\eqref{BWrelativistic} reproduces the results obtained
by the numerical integration of the complete formula
with an accuracy
better than $0.1\%$. 

\textcolor{black}{The resonant part of Eq.~\eqref{BWrelativistic}
is proportional to the $\Gamma_{ee}\Gh/\Gamma$ combination
which is a product of the partial width and the branching fraction:
$\Gamma_{ee}\times \tilde{\B}_h = \Gh\times \B_{ee}$.
Since our final result is $\Gamma_{ee}\times\B_h$,
let us consider the relation of $\Gh$ with the true
$\Gamma_h$, which is a sum of hadronic partial widths.}

\subsection{Interference effects in total multihadron cross section}
\label{Sec:Interference}
\textcolor{black}{
Considering charmonium decays at the parton level, one deals
with the gluonic $\psi\!\to\!gg(g/\gamma)$ and 
electromagnetic $\psi\!\to\!\gamma^{*}\!\to\!q\overline{q}$ modes.
\textcolor{black}{Treating quarks and gluons as final decay products,
one obtains that} 
gluonic modes do not interfere with the continuum 
$\ee\!\to\!\gamma^{*}\!\to\!q\overline{q}$ process while
those of electromagnetic origin do with the interference phase
equal to that of the dimuon decay.
In this case $\Gh$ does not differ from
a sum of the hadronic partial widths
$\Gamma_h = \Gamma_{gg(g\gamma)}+\Gamma_{q\overline{q}}\,$
with $\Gamma_{q\overline{q}}=R\,\Gamma_{ee}$
and
\begin{equation}
\lambda = \sqrt{\frac{R\,\mathcal{B}_{ee}}{\mathcal{B}_h}}\,,
\label{eq:lambda}
\end{equation}
where $\mathcal{B}_{ee}$ and $\mathcal{B}_h$ denote
the electron and hadron branching  fractions, respectively.
The real situation is much more complicated.
}

 For an exclusive hadronic mode $m$ 
\textcolor{black}{at a given point in the decay product phase space
$\Theta$} the amplitude $\ee\!\to\!m$ 
can be written as
\textcolor{black}{
\begin{equation}
\begin{split}
  A_m(\Theta) = & \sqrt{\frac{12\pi}{W^2}}\,\,
  \Bigg( \frac{\alpha}{3}\,\frac{a_m(\Theta)\sqrt{R_m}}{1\!-\!\Pi_0} \: - \\  
 & \frac{
   a_m(\Theta)\,M\sqrt{R_m\Gamma_{ee}^2}+
        a^{(s)}_m(\Theta)\,e^{i\phi_m}M\!\sqrt{\Gamma_{ee}\Gamma^{(s)}_m\,}
 }{M^2-W^2-iM\Gamma}
    \Bigg)\,,
\end{split}
\label{eq:Am}
\end{equation}}
where $R_m$ is the mode contribution to $R$, $\Gamma^{(s)}_m$
represents the contribution of the strong interaction to the
partial width  and
$\phi_m$ is its phase relative to the electromagnetic
contribution
$\Gamma^{(\gamma)}_m\!=R_m\,\Gamma_{ee}$\textcolor{black}{, the real
functions $a$ are normalized with $\int\!a_m^2(\Theta)\,d\Theta = 1$}.
In general, the phase $\phi_m$
depends on \textcolor{black}{$\Theta$}.
The numerator of the last term of Eq.\eqref{eq:Am} is proportional
to the decay amplitude, the partial width is
\begin{equation}
  \Gamma_m =  R_m \Gamma_{ee} + \Gamma^{(s)}_{m} +
    2\sqrt{R_m \Gamma_{ee} \Gamma^{(s)}_{m}\,}
            \left<\cos{\phi_m}\right>_{\Theta}\,,
\end{equation}
where the angle brackets
denote averaging over the \textcolor{black}{product phase space:
$\left<\,x(\Theta)\,\right>_{\Theta}\!\equiv\!
\int\!a(\Theta)a^{(s)}x(\Theta)\,d\Theta$}.
To obtain the exclusive mode cross section, the
following replacement must be done in
the expression \eqref{BWrelativistic}:
\begin{equation}
\begin{split}
& \Gamma_m \rightarrow \Gh\,, \:\:\:\:\:
   1 \rightarrow \lambda, \:\:\:\:\:
 R_m \rightarrow R\,, \\
& \Real{ \left(\!\sqrt{R_m \B_{ee}} \!+\!
     \left<e^{-i\phi_m}\right>_{\Theta}\!
          \sqrt{\frac{\Gamma^{(s)}_m}{\Gamma_m}}\,\right)
      \frac{f^{*}(W)}{1\!-\!\Pi_0}} 
      \,\rightarrow\,\Real{\frac{f^{*}(W)}{1\!-\!\Pi_0}} \,,
\end{split}
\end{equation}
where the latter replacement  follows from comparison of
the interference term corresponding to Eq.~\eqref{eq:Am}
and that of Eq.~\eqref{BWrelativistic}.
Performing them and summing over all hadronic modes one obtains the
expressions for $\Gh$ and $\lambda$:
\textcolor{black}{
\begin{equation}
\begin{split}
  \Gh =\, &\Gamma_h \,\,\times \\
    &\left(1+ 
     \frac{2\alpha}{3(1\!-\!\Re\Pi_0)\B_h} \sqrt{\frac{R}{\B_{ee}}}\,
   \!\sum\limits_m\!\sqrt{b_m \B^{(s)}_m}
       \left<\sin{\phi_m}\right>_{\Theta}
\right)
\label{eq:Gh}
\end{split}
\end{equation}
(here $\Gamma_h = \sum\limits_m\Gamma_m$, $\Im\Pi_0$ is neglected),
\begin{equation}
  \lambda = \sqrt{\frac{R \B_{ee}}{\B_h}} + \sqrt{\frac{1}{\B_h}}\,
        \sum\limits_m\!  
    \sqrt{b_m \B^{(s)}_m\,}
          \left<\cos{\phi_m}\right>_{\Theta}\,,
\label{eq:lambdaSum}
\end{equation}
where $b_m\!=\!R_m/R$ is a branching fraction of the continuum process
and $\B^{(s)}_m\!=\!\Gamma^{(s)}_m/\Gamma$.
Below, the sums
containing $\left<\sin{\phi_m}\right>_{\Theta}$ and 
$\left<\cos{\phi_m}\right>_{\Theta}$ 
are referred to as $\Sigma_{\sin}$ and $\Sigma_{\cos}$, respectively.
The parton level results are reproduced
by Eq.\eqref{eq:Gh} and~\eqref{eq:lambdaSum}
if $\Sigma_{\sin}$ and $\Sigma_{\cos}$
 can be neglected.} For a hypothetical
heavy charmonium decaying to light hadrons,
\textcolor{black}{both the values of
$\left<\cos{\phi_m}\right>_{\Theta}$ and
$\left<\sin{\phi_m}\right>_{\Theta}$ averaged over the product phase space
tend to zero}
due to the different \textcolor{black}{configuration of jets}
in electromagnetic
and strong decays. For the real $J/\psi$ and \PP one
has to rely on the absence of the phase correlations in different decays.
For the quasi-two-body decays such correlations are expected
(Ref.~\cite{PWang:interf} and references therein) but their branching
fractions are small~\cite{PDG}.

\textcolor{black}{If the sum $\Sigma_{sin}$
is not negligible, the method of the resonant cross section
determination employed in this and many other experiments
becomes inaccurate \textcolor{black}{and ambiguous}
because of the well-known ambiguity in the partial width
determination which takes place for each individual mode.
Indeed, a fit of the mode cross section $\sigma_m$ gives the values
of $\tilde{\Gamma}_m$ and $\cos{\phi_m}$ but leaves unknown
a sign of $\sin{\phi_m}$ required to obtain $\Gamma_m$.}

\textcolor{black}{Equating all $\left<\sin{\phi_m}\right>_{\Theta}$ in
Eq.~\eqref{eq:Gh} to unity, one sets an upper limit on
the inaccuracy of the hadronic partial width
\textcolor{black}{$\Delta\Gamma_h$} and 
\textcolor{black}{the
resonant cross section at the peak
\textcolor{black}{$\Delta\sigma^{\text{res}}_h(W_{\text{peak}})$}
used for the determination of the branching
fractions:
\begin{equation}
 \frac{\Delta\sigma^{\text{res}}_h(W_{\text{peak}})}
       {\sigma^{\text{res}}_h(W_{\text{peak}})} \approx
  \frac{\Delta\Gamma_h}{\Gamma_h} 
   \lesssim
     \frac{2\alpha}{3\B_h}\sqrt{\frac{R}{\B_{ee}}}\,
    \sum\limits_m\!\sqrt{b_m \B^{(s)}_m}\,.
\label{eq:GhInAccEst}
\end{equation} }
For \PP the sum in the right part $\lesssim 1-\B_{J/\psi+X}$,
thus Eq.~\eqref{eq:GhInAccEst} gives about 4\%. A better estimate
employing the $\lambda$ value obtained with the cross section fit 
is discussed in Section~\ref{err:inter}. Until this section we omit
the tilde mark wherever possible, thus $\Gamma_h$ and $ \B_h$ should be read as
\Gh and $\mathcal{\tilde{\B}}_h$.
}

\textcolor{black}{Eqs.~\eqref{eq:Gh} and ~\eqref{eq:lambdaSum} show that the
correct account of interference effects is essential for
a determination of the $\DP$ parameters due to 
its small value of $\B_{ee}$ and large
branching fraction to $D$ mesons\textcolor{black}{, nevertheless,
it was ignored in most of published analyses.
The interference effect is crucial for a
determination of the non--$D\overline{D}$ decay fraction of $\psi(3770)$
as was emphasized in Ref.~\cite{YangInterf}.
\SKIP{
The closeness to the $D\overline{D}$ threshold makes this even more
important because
of the asymmetry of the interference wave (Eq.~\eqref{BWrelativistic}
is not applicable near the threshold).}}} 

\subsection{Observed multihadron cross section}
\label{sec:OMHXS}

The multihadron cross section observed experimentally 
in the vicinity of \PP can be parameterized as follows:
\begin{equation}
\begin{split} 
\sigma_{\PP}^{\mathrm{obs}}(W) &= \epsilon_{\psi(2S)} \!\int\!
\sigma^{RC}_{\psi(2S)}(W^{\prime})G(W,W^{\prime})~\!dW^{\prime} \\
& + \epsilon_{\tau\tau}\,\sigma^{\tau\tau}_{\mathrm{cont}}(W) +
 \sigma_{\mathrm{cont}}(W)  \,.
\end{split}
\label{eq:fitpsi2s}
\end{equation}
Here $\epsilon_{\psi(2S)}$ and $\epsilon_{\tau\tau}$ are the detection
efficiencies {\textcolor{black}{and their dependence on $W$ can be
neglected}.
The continuum $\tau^{+}\tau^{-}$
cross section $\sigma_{\tau\tau}$
is included according to Ref.~\cite{Voloshin:tau} to extend
the validity of \eqref{eq:fitpsi2s} beyond the \PP-\DP region.
 
For the \PP cross section \eqref{BWrelativistic} includes the $\tau$
contribution and the $\lambda$ parameter is modified properly:
\textcolor{black}{
\begin{equation}
 \lambda_{h+\tau} \approx \sqrt{\frac{R\B_{ee}\,}{\B_h}}
+   \frac{\epsilon_{\tau\tau}}{\epsilon_{\PP}}\,
     \sqrt{\frac{R_{\tau}}{R}\frac{\B_{\tau\tau}}{\B_h}}
\label{eq:lamhtau}
\end{equation}
with $R_{\tau} =\sigma_{\tau^{+}\tau^{-}}/
\sigma_{\mu^{+}\mu^{-}}\approx 0.39$. The
reduction of $\tau^{+}\tau^{+}$ detection efficiency of about 0.3
compared to the multihadron one is accounted explicitly.
}

In Eq.~\eqref{eq:fitpsi2s} this cross section
is folded with the distribution over the total collision energy 
which is assumed to be quasi-Gaussian with an energy spread $\sigma_{W}$:
\begin{equation}
G(W,W^{\prime})=\frac{g(W\!-\!W^{\prime})}{\sqrt{2\pi}\sigma_{W}} 
\exp{\Bigg(-\frac{(W\!-\!W^{\prime})^2}{2\sigma_{W}^2}\Bigg)}~.  
\label{eq:QuasiGauss}
\end{equation}
\textcolor{black}{
The preexponential factor can be written as
\begin{equation}
g(\Delta) = \frac{1+a\,\Delta+b\,\Delta^2}{1+b\,\sigma_{W}^2}\,.
\label{eq:PreExp}
\end{equation}
It is due to various accelerator effects such as the
$\beta$-function chromaticity. 
We fix $a\!=\! b\!=\!0$ in our fit and consider the corresponding
systematic errors
\textcolor{black}{in the $\GGG\,$ product} in Section~\ref{err:detector}.
\textcolor{black}{The presence of this factor and other
accelerator- and detector-related effects yield too large systematic
uncertainties when the total width parameter $\Gamma$ is left floating in
the fit at $\sigma_W \gtrsim 5\times\Gamma/2$.
}}

\textcolor{black}{Since the interference effect is included
in $\sigma^{RC}_{\PP}$, the}
continuum cross section is a smooth function, which
with the sufficient accuracy can be parameterized with
\begin{equation}
\sigma_{cont}(W)=\sigma_{0}\, \Bigg(\frac{W_{0}}{W}\Bigg)^2\,,
\end{equation}
where $\sigma_{0}$ is the value of the background cross section at a
fixed energy $W_{0}$ below the \PP peak.

\textcolor{black}{In contrast with the commonly used interpretation
of the cross section as a sum of the resonant, continuum and interference
parts employed, in particular, in Ref.~\cite{BESelwid},
in the works~\cite{Ablikim:2006zq,Ablikim:2007gd,Ablikim:2008zzb}
it is interpreted as a sum of the two parts only:
the cross section of the resonance and the ``nonresonant'' one.
The latter is calculated using the full vacuum polarization
operator $\Pi_0\!+\!\Pi_{\PP}$ (Eq.~(10) of Ref.~\cite{Ablikim:2006zq}).
The two approaches
are equivalent provided that the bare parameters enter the
$\Pi_{\PP}$ and the electromagnetic contribution is excluded
from the cross section of the resonance (Eq.~(3) of
Ref.~\cite{Ablikim:2006zq}). The full vacuum polarization operator
describes not only the interference, but the electromagnetic decays
as well. If it is not done, the electron width extracted
from  the cross section fit would have a negative bias of about
$R\!\cdot\!\B_{ee}/\B_h$~$\approx0.018$.}

\subsection{Observed \ee cross section}

Bhabha scattering events detected in the calorimeter
were employed for luminosity measurements 
(see Section~\ref{sec:LumMeas} for more detail).
For the large angle Bhabha scattering
the  contribution of \PP decays is not negligible.
The differential \ee cross section
can be calculated with
\begin{equation}
  \begin{split}  
    &\left(\frac{d\sigma}{d\Omega}\right)^{ee\to ee} \!\!\!\approx 
    \left(\frac{d\sigma}{d\Omega}\right)_{\text{QED}}^{ee\to ee}+\\
    &\quad\frac{1}{M^2}
    \left\{\,\frac{9}{4}\frac{\Gamma^2_{ee}}{\Gamma
      M}(1+\cos^2\theta) \,\left(1\!+\!\frac{3}{4}\,\beta\right)\,\Im f -\right.\\
    &\qquad\left.\frac{3\alpha}{2}\frac{\Gamma_{ee}}{M}
    \left [(1+\cos^2\theta)-
      \frac{(1+\cos\theta)^2}{(1-\cos\theta)}\right ]
              \Re f
    \right\}\,.     
 \label{eq:eetoee}
\end{split}
\end{equation}
The first term represents the QED cross section calculated with
the Monte Carlo technique~\cite{BHWIDEGEN,MCGPJ}.
The second (resonance) and the third
(interference) terms have been obtained in~\cite{Azimov:eq}. The
corrections to the latter are not calculated
precisely, but that does not limit the accuracy of the published
results. Parameters $\beta$ and  $f$ are defined by
Eqs.~\eqref{eq:beta} and \eqref{resfunction}, respectively.

\section{Data analysis}
\subsection{Monte Carlo simulation}
\label{sec:MC}
The simulation of the experiment was performed 
in the frame of the {\it GEANT} package, version 3.21~\cite{GEANT:Tool}.

The \PP decays \textcolor{black}{and the continuum multihadron events}
were generated with the tuned version of the
BES generator~\cite{BESGEN} based on the \JETSET~7.4 code~\cite{JETSET}.
At the original parameter settings of the BES generator
the difference of the simulated charged multiplicity and that
observed experimentally exceeds $1\%$, thus the bias in the
detection efficiency up to $2\%$ is expected. 
The procedure of the parameter
tuning is discussed in detail below in Section~\ref{sec:mchadrerr}.
The decay tables were updated according to the recent PDG edition~\cite{PDG}.
The results are presented in Fig.~\ref{simdist_fig1}, where
the most important event characteristics obtained in the experiment
are compared with those in simulation.
Good agreement is observed.

The detection efficiency for $\tau^{+}\tau^{-}$ events was obtained using 
the KORALB event generator~\cite{KORALB24}.
Bhabha events required for the precise luminosity determination
were simulated using the BHWIDE generator~\cite{BHWIDEGEN}.
\begin{figure}[h]
\includegraphics[width=0.45\textwidth,height=0.2\textheight]{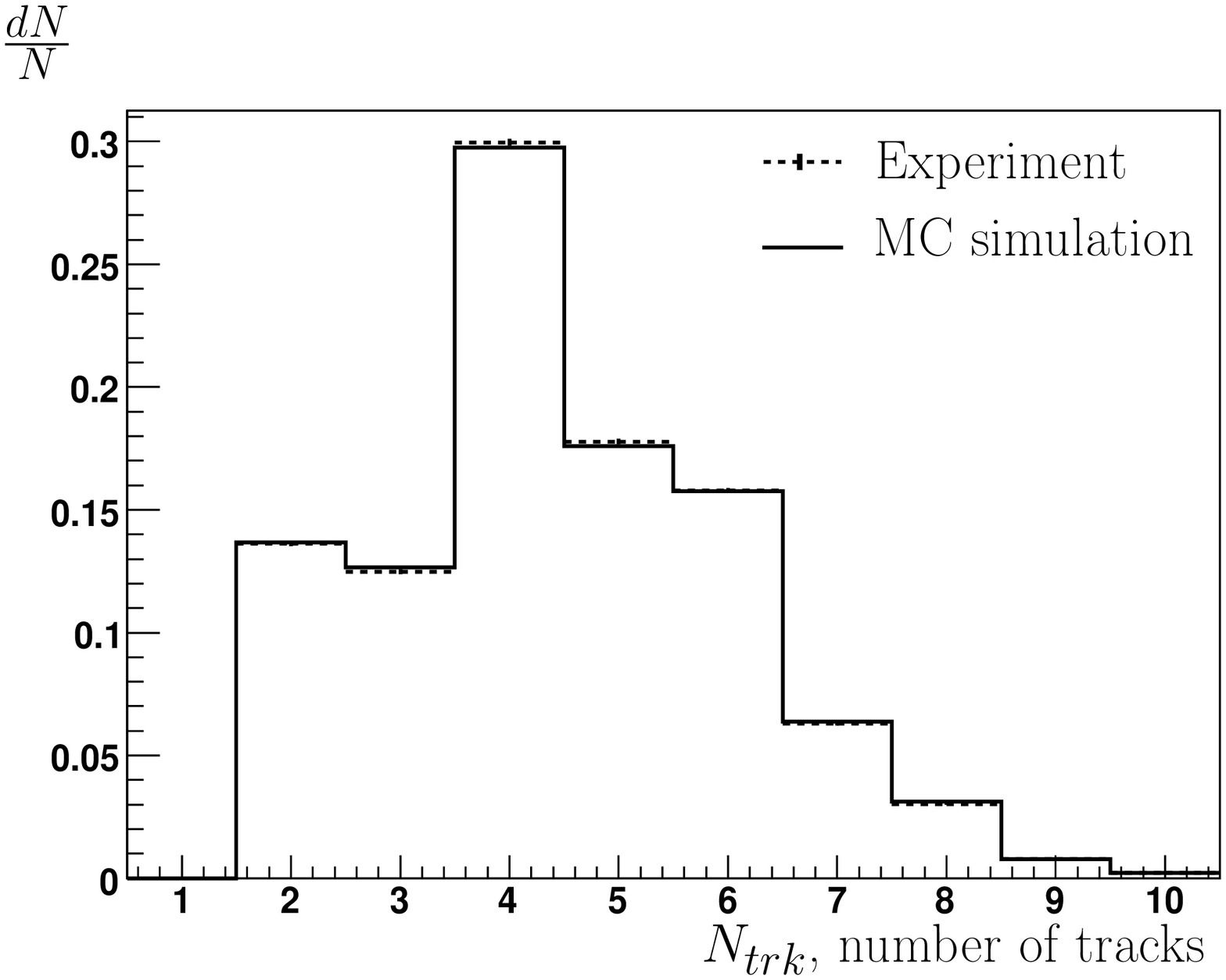}
\includegraphics[width=0.45\textwidth,height=0.2\textheight]{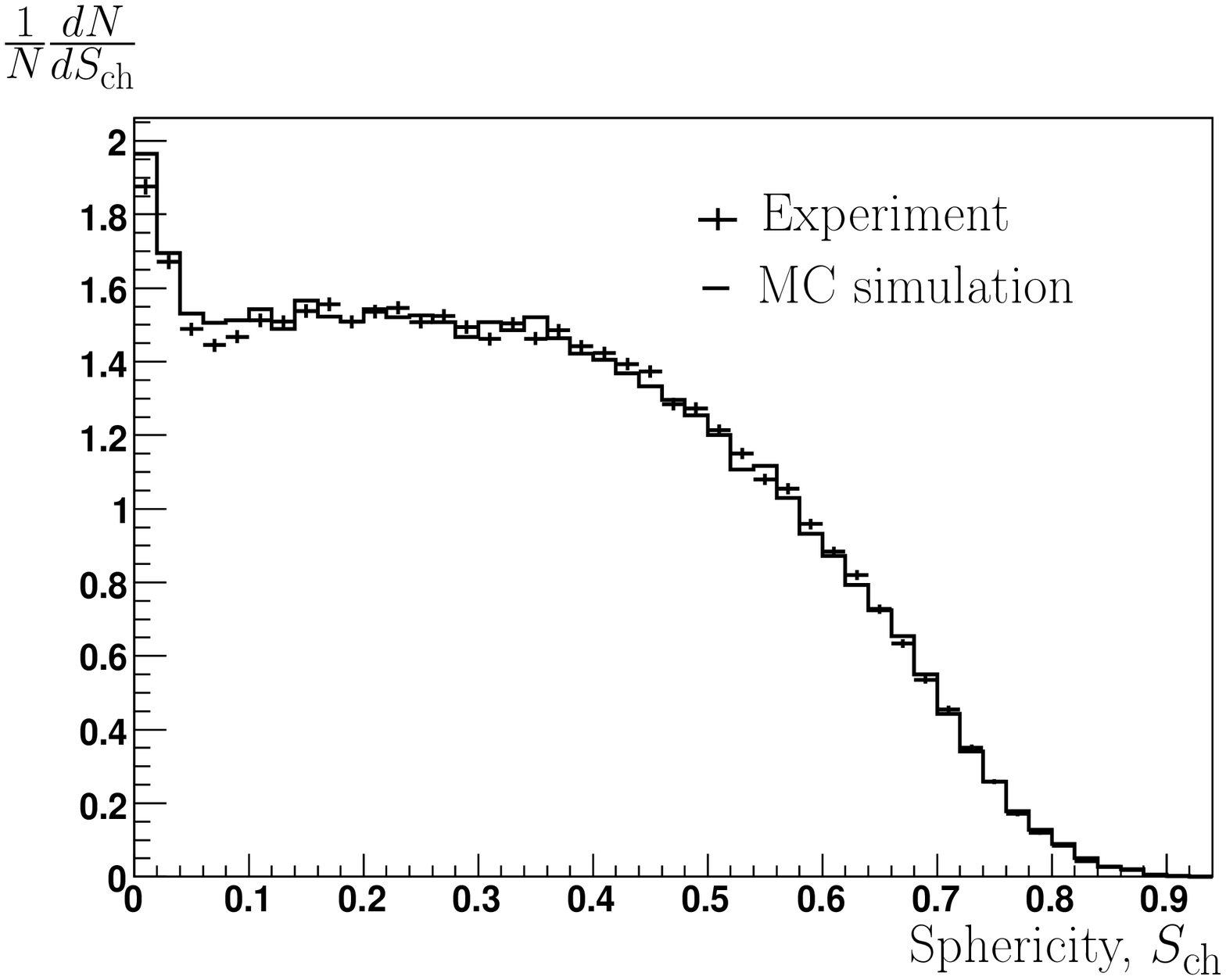}
\includegraphics[width=0.45\textwidth,height=0.2\textheight]{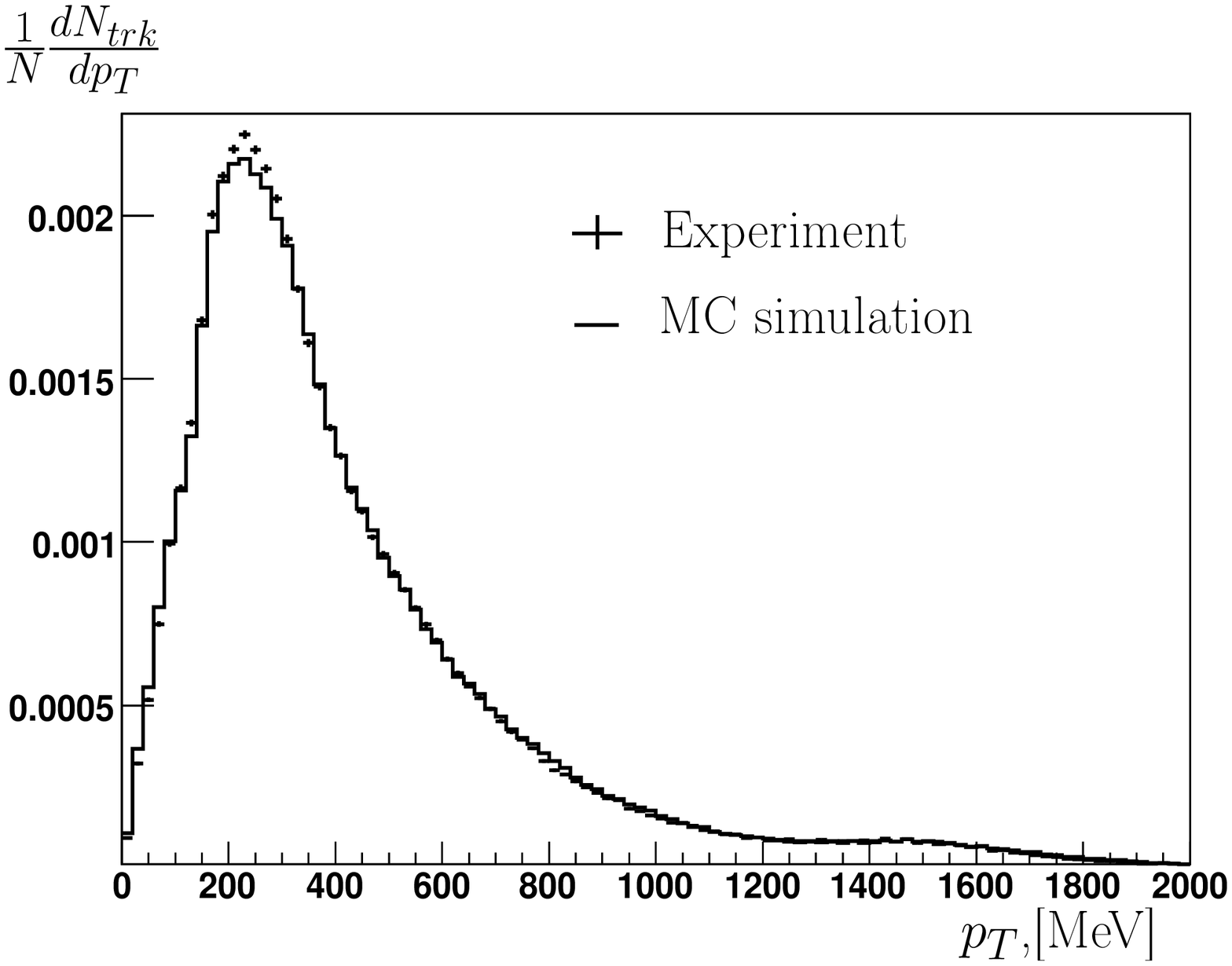}
\caption{{\normalsize Properties of hadronic events produced in \PP
    decay. Here $N$ is the number of events and 
    \textcolor{black}{
    $p_T$ is  the transverse momentum of a charged track}.
    All distributions are normalized to unity.} 
 }
\label{simdist_fig1}
\end{figure}
\subsection{Trigger efficiency and event selection}
\label{sec:trigger}
\begin{figure*}[t!]
\begin{center}
\includegraphics[width=0.67\textwidth]{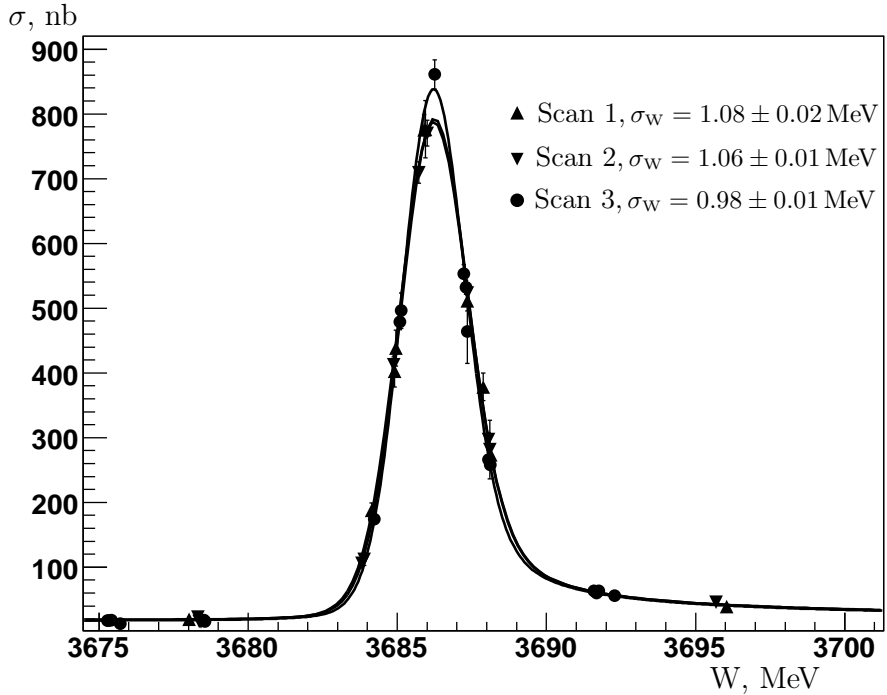}
\caption {\label{PsiPrimeScansFig}
{\normalsize
 The multihadron cross section as a function of the c.m. energy
 for three scans. The curves are the results of the fits.
 All data are corrected for the efficiency, \textcolor{black}{ the peak
 cross section depends on the energy spread $\sigma_W$.}
}}
\end{center}
\end{figure*} 
To reduce systematic errors due to trigger instabilities 
and uncertainties in the hardware thresholds,
 both experimental
and simulated events pass through the software event filter
during the offline analysis. It recalculates the PT and ST 
decisions with
tighter conditions using a digitized response of the detector
subsystems.
To suppress the machine background to an acceptable level, the
following PT conditions were used by OR:
\begin{itemize}\itemsep=-2pt
\item signals from $\ge$ 2 barrel scintillation counters\,,
\item signal from the LKr calorimeter in the scan of 2006\,, 
\item coinciding signals of two CsI endcaps.
\end{itemize}
Signals from two particles with the angular separation $\gtrsim\!20$
degrees satisfy
the ST conditions which are rather complicated.
The MC simulation yields the trigger efficiency of about 0.96 for \PP decays.
Because of a problem with electronics, the LKr calorimeter 
was not used in the analysis of 2004 data and that decreased the 
trigger efficiency to 0.91. 

The performance of the detector subsystems and the machine
background conditions were very different in 2004 and 2006,
so that the selection criteria are also different.
\begin{center}
  2004 data, first and second scans:
\end{center}
\vspace*{-1.75ex}
\begin{itemize}\itemsep=-2pt
\item  $\geq$ 3 charged tracks\,,
\item  $\geq$ 2 charged tracks from a common vertex in the interaction
region (\mbox{$\,\rho\!<\!7$}~mm,\, \mbox{$|z|\!<\!130$}~mm)\,,
\item event sphericity $S_{\!\text{ch}}\!>\!0.05$\,.  
\end{itemize}
Here $\rho$ and $z$ are the track impact parameters
relative to the beam axis
and $z$-coordinate of the closest approach point.

\textcolor{black}{
The sphericity parameter is defined as
\begin{equation}
   S = \frac{3}{2} \min\frac{\sum p_{T,i}^2}{\sum p_i^2}\,,
\end{equation}
where summation is performed over all particles of the event
and the minimum is taken over directions of the axis relative
to which the transverse momenta $p_{T,i}$
are calculated.
$S_{\!\text{ch}}$ is calculated using charged tracks only.
The cut on $S_{\!\text{ch}}$}  
is efficient for suppression of the $\ee\!\to\!\ee\gamma\,$
background\textcolor{black}{, that of cosmic rays and some kinds of the
machine background}, though it also suppresses the leptonic modes of the
cascade decay $\PP\!\to\!J/\psi\!+\!\text{{\it neutrals}}$
(\textcolor{black}{see the low sphericity peak in} Fig.~\ref{simdist_fig1}).
\begin{center}
  2006 data, third scan:
\end{center}
\vspace*{-1.75ex}
\begin{itemize}\itemsep=-2pt
\item  $\geq$ 3 charged tracks or two tracks with the acollinearity
 $>$~35~degrees,
\item  $\geq$ 2 charged tracks from a common vertex in the interaction
region (\mbox{$\,\rho\!<\!7$}~mm,\, \mbox{$|z|\!<\!130$}~mm)\,,
\item  $\geq$ 1 photons with energy $\geq$~100~MeV in the
calorimeter\,,
\item event sphericity $S_{\!\text{ch}}\!>\!0.05$\,.
\item  the energy deposited in the calorimeter $\geq$ 450 MeV\,. 
\end{itemize}

Analyzing the third scan we also used  the alternative selection criteria
without a tight cut on the sphericity, but with additional requirements
on the calorimeter response.
It allows us to check the systematic error due to the sphericity
cut.

\textcolor{black}{For additional suppression of the background induced
by cosmic rays a veto from the muon system was required in the cases
when more than two tracks did not cross the interaction region or
the event arrival time determined by TOF relative to the bunch crossing
was less than -5 ns or larger than 10 ns. This condition was common
for all three scans.
}

The conditions described above reduce the physical background contributions
which do not scale with energy like $1/s$
to a negligible level\textcolor{black}{, except the tau pair production
which we took into account explicitly in the observed cross section.
The contribution of the beam--wall and beam-gas events, cosmic events
and their coincidences was evaluated using
data collected with the separated beams (about 10\% of the
full data sample). It was about
2\% of the observed continuum cross section for the third scan
and about 0.4\% for the first two. The analysis of the
event vertex distribution along the beam axis confirmed these
estimates. We did not perform the background subtraction in each data
point as was done in the work~\cite{BES3pnt} for precise $R$ measurements.
This is not required for the resonance parameter determination, the
corresponding uncertainties are discussed in
Section~\ref{err:detector}.}
Simulation of \PP decays yields  the detection efficiencies of
0.63 and 0.72 for
the two sets of selection criteria, respectively.
To ensure the detection efficiency stability, all 
electronic channels malfunctioning in some runs during a scan were excluded 
from the analysis of all its runs.

\subsection{Luminosity determination} \label{sec:LumMeas}

The stability and absolute calibration accuracy of the \\
bremsstrahlung monitors
used for on-line luminosity measurements (Section~\ref{sec:VEPP})
are not sufficient for the
precision cross section analysis,
thus events of Bhabha scattering 
were employed for the off-line luminosity determination.
In 2004 it was provided by the endcap CsI calorimeter (the fiducial region
$20^{\circ}\!<\!\theta\!<\!32^{\circ}$ 
\textcolor{black}{and $148^{\circ}\!<\!\theta\!<\!160^{\circ}$}). 
In the analysis of the 2006 data the 
LKr calorimeter was employed ($40^{\circ}\!<\!\theta\!<\!140^{\circ}$) 
while the CsI one served for cross-checks only.

The criteria for \ee event selection using the calorimeter data are
listed below:
\textcolor{black}{
\begin{itemize}\itemsep=-2pt
\item Two clusters with
      the energy above 0.25 of the beam energy and the angle
      between them exceeding 165 degrees,
\item The total energy of these two clusters exceeds 1.05 times
      the beam energy,
\item The calorimeter energy not associated with
      these two clusters does not exceed 10$\%$ of the
      total.
\end{itemize}
The loose energy cuts were chosen to reduce the influence of the
calorimeter channels excluded from the analysis as was mentioned
above.}
The tracking system was used only to reject
the background ($\ee\!\to\!\gamma\gamma$ and $\ee\!\to\!\text{hadrons}$).
The number of extra photons was required to be less than two for the
additional suppression of the latter.

\subsection{Fitting procedure} 
\label{sec:fit}

The collision energy $W$ was assigned to each data acquisition run
using the interpolated results of the beam energy measurements
and assuming
$W\!=\!2E_{beam}$. The runs with
close $W$ values were joined into points with the luminosity-weighted
values $W_{i}$ ($i$ is the point number). 

The numbers of hadronic events $N_{i}$ and
events of Bhabha scattering $n_{i}$ observed at the $i$-th energy point
were fitted as a function of $W$ using the
maximum likelihood method 
with a likelihood function
\begin{equation}
\begin{split}
-2 \,\ln{ \mathcal{L}} & = 2 \sum_{i} \Bigg[ N_{i}^{\mathrm{obs}} 
  \ln{ \Bigg( \frac{N_{i}^{\mathrm{obs}}}{ N_{i}^{\mathrm{exp}} } \Bigg)} +  N_{i}^{\mathrm{exp}} -  N_{i}^{\mathrm{obs}} \\
  +& n_{i}^{\mathrm{obs}} \ln {\Bigg( \frac{ 
  n_{i}^{\mathrm{obs}} }{ n_{i}^{\mathrm{exp}} } \Bigg)}+
  n_{i}^{\mathrm{exp}} -  n_{i}^{\mathrm{obs}} \Bigg]   \,,
\label{ML:func}
\end{split}
\end{equation}
where  $N_{i}^{\mathrm{exp (obs)}}$ and $\,n_{i}^{\mathrm{exp(obs)}}$ are the expected
(observed) numbers of the hadronic and Bhabha events, respectively.
The expected numbers of the hadronic and  Bhabha
scattering events were determined as follows: 
\begin{equation}
\begin{split}
 N_{i}^{\mathrm{exp}} & = \sigma_{hadr}(W_i) \cdot L_{i}\,,\\
 n_{i}^{\mathrm{exp}} & = \sigma_{\ee}(W_i) \cdot L_{i}\,,
\end{split}
\end{equation}
here  $\sigma_{hadr}$ and $\sigma_{\ee}$ are defined by
\begin{equation}
\begin{split}
\sigma_{hadr} & = \sigma^{\mathrm{obs}}_{\PP}(W_i)+
\epsilon_{\ee(hadr)}\,\sigma_{\ee}^{\mathrm{obs}}(W_i)\,, \\
\sigma_{\ee} & =\sigma_{\ee}^{\mathrm{obs}}(W_i) +   \epsilon_{hadr(\ee)}\, \sigma^{\mathrm{obs}}_{\PP}(W_i),
\end{split}
\end{equation}
where $\sigma^{\mathrm{obs}}_{\PP}(W_i)$
and $\sigma_{\ee}^{\mathrm{obs}}(W_i)$ 
were calculated according to \eqref{eq:fitpsi2s} and by
integration of \eqref{eq:eetoee},
respectively.
The detection efficiencies entering the formulae
were determined separately at each point using the run-dependent
Monte Carlo simulation.
The values of the cross-feed selection efficiency $\epsilon_{\ee(hadr)}$
(the probability to select the $\ee\!\to\!\ee$ event
as the hadronic one) obtained by MC are about $0.006\%$,
$0.006\%$ and $0.37\%$ for three scans, respectively.
The corresponding values of $\epsilon_{hadr(\ee)}$
are $0.03\%$, $0.03\%$ and $0.25\%$.

The integrated luminosity $L_{i}$
at the energy point $i$ can be derived from the condition
\( \partial/\partial L_{i}\ln\mathcal{L} = 0~, \)
giving
\begin{equation}
L_{i}  = \frac{N_{i}^{\mathrm{obs}}+n_{i}^{\mathrm{obs}}}{\sigma_{hadr}+
  \sigma_{\ee}}\,.
\label{Lum:def}
\end{equation}
Using the likelihood function that takes into account both $N_i$ and $n_i$ 
ensures a correct estimation of the statistical uncertainty
in the fit results.

The data of each scan were fitted separately, 
the free parameters were the \PP
mass $M$, $\GGG$, the energy spread $\sigma_{W}$ and the continuum
cross section magnitude $\sigma_0$. \textcolor{black}{The $\lambda$ parameter
was fixed at the value of 0.13 according to Eq.\eqref{eq:lamhtau}.}
The data points ($\sigma_{i},W_{i}$) and the fitted curves are
shown in~Fig.~\ref{PsiPrimeScansFig}.
The results of the fits are presented
in Table~\ref{tab:psi2S}. 
\begin{table}[t!]
\vspace*{-2ex}
\begin{center}
\caption {\label{tab:psi2S} {\normalsize The main results of the scan fits 
(statistical errors only are presented).}}
\begin{tabular}[l]{|l|c|c|c|} \hline
               &$M\,,\,\text{MeV}$ & $\GGG\,,\, \text{keV}$ &
$P(\chi^2)$, $\%$\\\hline
Scan 1   &  $3686.102 \pm 0.018$   & $2.258 \pm 0.033$ &15.8\\ 
Scan 2   &  $3686.130 \pm 0.013$   & $2.229 \pm 0.024$ &29.5\\  
Scan 3   &  $3686.108 \pm 0.010$   & $2.226 \pm 0.022$ &79.5\\\hline
\end{tabular}
\end{center}
\end{table}

The statistical accuracy of the mass values is
significantly better than that for \textcolor{black}{the}
three scans performed at VEPP-4M in 2002~\cite{Aulchenko:2003qq}.
\textcolor{black}{The combined analysis is required
to take into account properly the systematic errors
of the six mass values. It will be described in a separate paper
discussing numerous accelerator-related effects. For this reason
we omit such a discussion below and just present the result
and error estimates. \textcolor{black}{The analysis of
the three values obtained in this work gives
\begin{equation*}
    M = 3686.114 \pm 0.007 \pm 0.011 \:\,^{+0.002}_{-0.012} \:\:\text{MeV}.
\end{equation*} }
The model dependence of the mass value
was estimated floating the $\lambda$ parameter in the fit.}

\section{Discussion of systematic uncertainties in \GGG}
\label{sec:Errors}
The dominating sources of the systematic uncertainty in the
\textcolor{black}{$\Gamma_{ee}\times\tilde{\B}_h = \Gamma_{ee}\times\Gh/\Gamma$
value}  are discussed in the following subsections. \textcolor{black}{
The issue of the difference between $\Gh$ and the sum of
hadronic partial widths because of possible correlations
of interference phases is addressed in the next section.}

\subsection{Systematic error of \textcolor{black}{absolute} luminosity
determination}
\label{sec:lumerr}

The major contributions to the uncertainty of the
\textcolor{black}{absolute} luminosity determination are
presented in Table~\ref{tab:lumerr}.

\begin{table}[h!]
\vspace*{-2ex}
\caption{{\normalsize \label{tab:lumerr} Systematic uncertainties of
the luminosity determination in \% for  three scans.
The correlated parts of the uncertainties are also presented.
The uncertainties for the first and second scans
are assumed to be fully correlated.
}}
\begin{center}
\begin{tabular}{lllll}
\emph{Source} &
         \multicolumn{1}{c}{\hspace*{-0.25em}Scan 1} &
         \multicolumn{1}{c}{\hspace*{-0.5em}Scan 2} & 
         \multicolumn{1}{c}{\hspace*{-0.5em}Scan 3} & 
             \multicolumn{1}{l}{\hspace*{-0.75em}Common} \\*[2pt]
\hline
Calorimeter calibration\hspace*{-1em}
                        &\BL  0.3  &\BL  0.3 &\BL 0.7  &\BL 0.2 \TNL \hline
Calorimeter alignment   &\BL  0.3  &\BL  0.3 &\BL 0.1  &\BL 0.1 \TNL\hline
Polar angle resolution
                        &\BL  0.8  &\BL  0.8 &\BL 0.2   &\BL - \TNL\hline
Cross section calculation\hspace*{-1em}
                        &\BL  0.5  &\BL  0.5 &\BL 0.5  &\BL 0.4  \TNL\hline
Background              &\BL  0.1  &\BL  0.1 &\BL 0.1  &\BL 0.1 \TNL\hline
MC statistics           &\BL  0.1  &\BL  0.1 &\BL 0.1  &\BL -  \\ \hline
Variation of cuts       &\LT  1.3  &\LT  1.4 &\LT 0.9  &\LT 0.2         \TNL\hline\hline
\emph{Sum in quadrature} &\AP  1.6 &\AP1.7 & \AP  1.2     &\AP 0.5 \\
\end{tabular}
\end{center}
\end{table}

The uncertainty due to the calorimeter energy calibration was estimated
by variation of simulation parameters
concerning the GEANT performance, the light yield uniformity
in CsI crystals, a sensitivity to the energy loss fluctuations between
LKr calorimeter electrodes etc. Additionally, the CsI individual
channel calibrations using cosmic rays and  Bhabha events were compared.

The LKr calorimeter was aligned with respect to the drift chamber using
cosmic tracks reconstructed in the DC.
Then eight modules of the CsI calorimeters were
aligned relative to the LKr calorimeter using cosmic rays reconstructed
in the LKr strip system. The beam line position and direction were determined
using the primary vertex distribution of multihadron events. The inaccuracy
of the alignment resulted in the error of the luminosity measurements
of about $0.3\%$ for CsI and about $0.1\%$ for LKr.

The difference in the polar angle resolutions observed experimentally
and predicted by MC
causes an uncertainty in the luminosity measurement, since events migrate 
into or out of the fiducial volume.
These errors are $0.8\%$ and $0.2\%$ for the CsI and LKr calorimeters,
respectively.

The uncertainty of the theoretical Bhabha cross section
was estimated comparing the results obtained with
the BHWIDE~\cite{BHWIDEGEN} and MCGPJ~\cite{MCGPJ} event generators.
It agrees with the errors quoted by the authors. \textcolor{black}{
The value of 0.5\% quoted in Table~\ref{tab:lumerr}
includes also the accuracy of Eq.~\eqref{eq:eetoee}.}

The background to the Bhabha process  from the 
$\psi(2S)$ decays and reactions $\ee \to \mu\mu(\gamma)$ and $\ee \to \gamma
\gamma $ was estimated using MC simulation.
It contributes  less than $0.3\%$ to the luminosity. 
The resonant part of the background contribution 
was taken into account in the fit
(Section~\ref{sec:fit}). The residual luminosity
uncertainty due to background  does not exceed $0.1\%$.

In order to estimate the effect of other possible sources of uncertainty,
the variation of the cuts was performed within the fiducial
region in which good agreement between the MC simulation and 
experiment is observed.
The cut on the deposited energy was varied in the range of $55-75\%$
of the c.m. energy. 
The cuts on the polar angle were varied in a range much larger than the
angular resolution, the variation in the Bhabha event count
reaches $40\%$. 
The variations discussed above correspond to a systematic uncertainty
shown in Table~\ref{tab:lumerr}. 
These effects can originate from the already considered
sources and statistical fluctuations, nevertheless we included
them in the total uncertainty to obtain conservative error estimates.

Finally, we compared an integrated luminosity
obtained using the LKr and CsI calorimeters in the scan of 2006.
The difference of about $1.1\pm 1.0\%$ was found which
is consistent with the estimates in Table~\ref{tab:lumerr}.

\subsection{Uncertainty due to imperfect simulation of \PP decays}
\label{sec:mchadrerr}

\begin{table*}[t!]
\caption{{\normalsize\label{err:mchadrtable} Comparison of different
versions of the MC simulation of \PP decays.}}
\begin{center}
\begin{tabular}[l]{|c|c|c|c|c|c|l|l|l|} \hline
Version  & \multicolumn{6}{|c|}{ \JETSET\ modifications }   &Charged multiplicity  & Efficiency, $\%$           \\\hline  
\multicolumn{9}{|c|}{ {\bf LUND fragmentation function} } \\\hline 
\multicolumn{9}{|c|}{{\it  \textcolor{black}{Probability of vector meson formation}}}\\\hline 
    & \multicolumn{3}{|c|}{ $P_{ \mathrm{V} }$}  &    \multicolumn{3}{|c|}{$\sigma_{p_T}\,,\,\text{GeV}$ }  &  &           \\\hline  
1 &  \multicolumn{3}{|c|}{0.50} & \multicolumn{3}{|c|}{0.55}       &$4.1391\pm 0.0031$& $71.392 \pm0.090$ \\\hline  
2 &  \multicolumn{3}{|c|}{0.50} & \multicolumn{3}{|c|}{0.65}       &$4.1555\pm 0.0031$& $72.232 \pm0.090$  \\\hline  
\multicolumn{9}{|c|}{{\it $W_{\mathrm{stop}},\,W_{\mathrm{min}},\,\sigma_{p_T}$ varied  }} \\\hline 
        & \multicolumn{2}{|c|}{$W_{\mathrm{stop}}\,,\,\text{GeV}$}        &\multicolumn{2}{|c|}{$W_{\mathrm{min}}\,\text{GeV}$}  &       \multicolumn{2}{|c|}{$\sigma_{p_T}\,,\,\text{GeV}$} &     &    \\\hline  
6 & \multicolumn{2}{|c|}{0.47} &\multicolumn{2}{|c|}{0.8}&\multicolumn{2}{|c|}{0.65} &   $4.1429\pm0.0031$ & $71.783\pm0.090$   \\\hline  
7 & \multicolumn{2}{|c|}{0.52} &\multicolumn{2}{|c|}{0.8}  & \multicolumn{2}{|c|}{0.65} &  $4.1488\pm0.0031$& $72.049 \pm0.090$  \\\hline 
8 & \multicolumn{2}{|c|}{0.56} &\multicolumn{2}{|c|}{0.8}  & \multicolumn{2}{|c|}{0.65} &  $4.1512\pm0.0031$ & $72.170\pm0.090$   \\\hline  
\multicolumn{9}{|c|}{{\it $\delta W_{\mathrm{stop}},\,\sigma_{p_T}$ varied }} \\\hline 
   & \multicolumn{3}{|c|}{ $\delta W_{\mathrm{stop}}$}  &  \multicolumn{3}{|c|}{$\sigma_{p_T}\,,\,\text{GeV}$}   &  &           \\\hline  
9  &  \multicolumn{3}{|c|}{0.17} & \multicolumn{3}{|c|}{0.7}       & $4.1407\pm 0.0031$& $71.890 \pm0.090$ \\\hline  
10 &  \multicolumn{3}{|c|}{0.17} & \multicolumn{3}{|c|}{0.65}      & $4.1552\pm 0.0031$& $72.430 \pm0.090$  \\\hline  
\multicolumn{9}{|c|}{{\it $W_{\mathrm{min}},\sigma_{p_T}$    varied }} \\\hline 
  & \multicolumn{3}{|c|}{ $W_{\mathrm{min}}\,,\,\text{GeV}$ }  &  \multicolumn{3}{|c|}{$\sigma_{p_T}\,,\,\text{GeV}$}   &  &       \\\hline  
11  &  \multicolumn{3}{|c|}{0.8} & \multicolumn{3}{|c|}{0.675}      &$4.1401\pm 0.0031$& $72.033 \pm0.090$   \\\hline  
12 &  \multicolumn{3}{|c|}{0.8}  & \multicolumn{3}{|c|}{0.65}       &$4.1529\pm 0.0031$& $72.378 \pm0.090$  \\\hline   
\multicolumn{9}{|c|}{ {\bf Switched off parton shower} } \\\hline 
      & \multicolumn{6}{|c|}{ $\sigma_{p_T}\,,\,\text{GeV}$ }  &    &           \\\hline  
13    &  \multicolumn{6}{|c|}{ 0.65 }  &  $4.1409\pm 0.0031$ &   $72.118\pm0.090$         \\\hline  
14    &  \multicolumn{6}{|c|}{ 0.55 }  &  $4.1554\pm 0.0031$  &  $72.709 \pm0.090$        \\\hline  
\multicolumn{9}{|c|}{{\bf Field-Feynman fragmentation function}} \\\hline 
       & \multicolumn{3}{|c|}{$W_{\mathrm{stop}}\,,\,\text{GeV}$}  &  \multicolumn{3}{|c|}{$\sigma_{p_T}\,,\,\text{GeV}$} &     &    \\\hline  
3      & \multicolumn{3}{|c|}{0.62}    &  \multicolumn{3}{|c|}{0.58} &$4.1372\pm0.0031$ &  $71.475\pm0.090$   \\\hline   
4      & \multicolumn{3}{|c|}{0.62 }   &  \multicolumn{3}{|c|}{0.50} &$4.1491\pm0.0031$ &  $71.981\pm0.090$   \\\hline  
5      & \multicolumn{3}{|c|}{0.62 }   &  \multicolumn{3}{|c|}{0.43} &$4.1650\pm0.0031$ &  $72.755\pm0.090$    \\\hline  
\end{tabular}
\end{center}
\end{table*}

The imperfect simulation of \PP decays contributes
significantly to the \GGG\, systematic error caused by
the detection efficiency uncertainty. Eventually, this uncertainty
is determined by the experimental statistics available for the event
generator tuning and the ability of the latter to reproduce
distributions of parameters essential for the event selection
and their correlations.
In our case such parameters are the charged multiplicity and
event sphericity calculated using momenta of charged tracks.

The selection criteria described in Section~\ref{sec:trigger}
reject low-multiplicity events. The corresponding
branching fractions are either negligible like for $\PP\!\to\!\pi^{+}\pi^{-}$
or well measured like for 
$\PP\!\to\!J/\psi\,\eta\!\to\!\mu^{+}\mu^{-}\gamma\gamma\,$.
Such decays are simulated in the BES generator~\cite{BESGEN} explicitly.
The variation of decay table parameters within their errors (Ref.~\cite{PDG})
indicates the detection efficiency uncertainty of less than or about 0.3\%.
The dominating uncertainty comes from decays of higher multiplicity
which are simulated using the parton approach
and the fragmentation model
incorporated in the \JETSET~7.4 code.

About 60\% of the \PP decays include \JP which is at the edge or
even beyond the region where the \JETSET\ results can be trusted, nevertheless
it has enough options and parameters to achieve good agreement
with experiment for major event characteristics
(Fig.~\ref{simdist_fig1}) and to estimate
the detection efficiency uncertainty. To do this we iterated as follows:
\begin{enumerate}
\item select a critical option or parameter and modify it using an
educated guess;
\item select a complementary parameter and modify it to find
the value at which the observed
charged multiplicity agrees with experiment;
\item calculate the detection efficiency and compare it with
previous results to estimate the uncertainty.
\end{enumerate}
In addition to the charged multiplicity, the observed distributions
in the sphericity parameter, the invariant mass of the pairs of the opposite
signs and the inclusive momentum spectrum were controlled. The versions
of the simulation obviously contradicting to experiment were rejected.
The results are presented in Table~\ref{err:mchadrtable} and illustrated by
Fig.~\ref{err:mchadrfig}.
\begin{figure}[h!]
\includegraphics[width=0.5\textwidth]{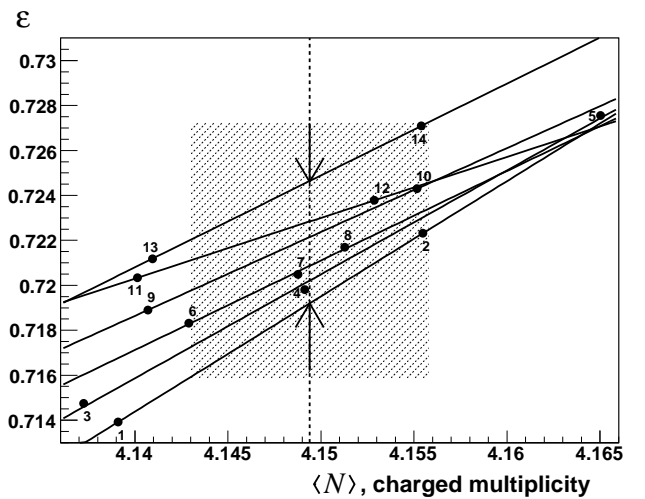}
\caption{{\normalsize \label{err:mchadrfig}Detection efficiency vs.$\!$
charged multiplicity for different versions of the \PP
decay simulation. The \textcolor{black}{fit} lines correspond to variation of one of the
selected parameters. The shadow box corresponds to the statistical
error
of the charged multiplicity. The statistical error of the efficiency
$\sim\,$0.001 is not shown.}}
\end{figure}

The following \JETSET\ options were studied:
\begin{enumerate}
\item LUND fragmentation function, parton showers are on;
\item LUND fragmentation function, parton showers are off;
\item Field-Feynman fragmentation function, parton showers are on;
\item independent fragmentation, all momentum conservation options.
\end{enumerate}
No acceptable versions were obtained in the latter case. For the LUND
fragmentation function the influence of a few most critical parameters
was investigated.

 The main parameters controlling the fragmentation process in
\JETSET\ are PARJ(32), PARJ(33) and PARJ(37) (we refer to them as
$W_{\mathrm{min}}$, $W_{\mathrm{stop}}$ and $\delta W_{\mathrm{stop}}$, 
respectively, in Table~\ref{err:mchadrtable} and below). The fragmentation
of a colour-singlet system formed by initial partons or during
the parton showering proceeds while the energy is greater than
$W_{\mathrm{min}}$ plus quark masses, otherwise 
a pair of hadrons is to be produced.
The $W_{\mathrm{stop}}$ parameter serves to terminate
the fragmentation and produce a final hadron pair
earlier, taking into account the mass of the
last quark pair produced. To avoid artifacts in the hadron
momentum spectrum, the stopping point energy is smeared using
the $\delta W_{\mathrm{stop}}$ parameter (20\% by default).
The transverse momentum of quarks appearing during the fragmentation
is controlled by the parameter PARJ(21) ($\sigma_{p_T}$).
The parameter PARJ(11) ($P_{V}$) determines the probability
that a light meson formed in the fragmentation has spin 1.
The default values of the parameters $W_{\mathrm{min}}$, $W_{\mathrm{stop}}$,
$\sigma_{p_T}$ and $P_{V}$ in \JETSET\ 
are 1., 0.8, 0.36 GeV and 0.5, respectively.
The corresponding values set in the BES generator are 1., 0.6, 0.5 GeV
and 0.6.


Figure~\ref{err:mchadrfig} shows the values of the detection
efficiency and observed charged multiplicity for 
the sets of options and parameters listed in Table~\ref{err:mchadrtable}.
The dashed line corresponds to the charged multiplicity observed 
in experiment with the selection criteria of the 2006 scan,
$\langle N_{\mathrm{exp.}} \rangle = 4.1494 \pm 0.0054$.
The maximum difference between the detection efficiency values
at the point of the multiplicity agreement is $0.55\%$ 
\textcolor{black}{with the central value of about 72.19\%. This is 
presented with the segment between the arrows and its middle point
in Fig.~\ref{err:mchadrfig}.
The experimental and simulated multiplicities have statistical errors,
therefore the segment of the agreement transforms to the rectangle
shown with the shadow box. Taking its angles we found the 
confidence interval of $(72.16 \pm 0.57)$\% and from that derived that
the relative uncertainty of the detection efficiency due to ambiguity in the 
choice of the \JETSET\ parameter set is about $0.8\%$.
A very similar central value of 72.11\% was obtained by averaging the
fourteen efficiencies from Table~\ref{err:mchadrtable} with the weights
inversely proportional to the sum of $\chi^2$ for the four event
characteristics under control.
}

There is a systematic error in the observed multiplicity 
related to the track reconstruction efficiency, which
is not exactly the same for the experimental data and
simulation. The difference was studied using
Bhabha events and low-momentum cosmic tracks and the
appropriate correction was introduced 
in the detector simulation. However, the inaccuracy of
the correction increases the shadow box size in
Fig.~\ref{err:mchadrfig} thus increasing the detection efficiency
uncertainty. For the first two scans the effect is about 0.4\% and
it grows up to 0.7\% for the third scan because of some problems
in the drift chamber.
\begin{table*}[t]
\caption{{\normalsize\label{tab:mcerr} Systematic uncertainties of the
detection efficiency due to \PP decay simulation
in \% for three scans.
The correlated parts of the uncertainties are also presented.
The uncertainties for the first and second scans
are assumed to be fully correlated.}}
\begin{center}
\begin{tabular}{lllll}
\emph{Source}
      & \multicolumn{1}{c}{\hspace*{-0.5em}Scan 1} &
      \multicolumn{1}{c}{\hspace*{-0.5em}Scan 2} & 
      \multicolumn{1}{c}{\hspace*{-0.5em}Scan 3} & 
      \multicolumn{1}{l}{\hspace*{-0.75em}Common} \\*[2pt]
\hline
Measured branchings       
               &\BL  0.4   &\BL  0.4       &\BL  0.3&\BL 0.3 \TNL\hline
\JETSET\ ambiguities
               &\BL  0.8   &\BL  0.8       &\BL  0.8&\BL 0.8 \TNL\hline
Track reconstruction
               &\BL  0.4   &\BL  0.4       &\BL  0.7&\BL 0.4 \TNL\hline
Selection criteria
               &\BL  0.3   &\BL  0.3       &\BL  0.3&\BL 0.3 \TNL\hline
MC statistics
               &\BL  0.1   &\BL  0.1       &\BL  0.1& \BL -\\ \hline\hline
\emph{Sum in quadrature}
               & \AP 1.0 &\AP    1.0       &\AP 1.1 &\AP 1.0 \\
\end{tabular}
\end{center}
\end{table*}
\begin{table*}[t!]
\caption{{\normalsize\label{tab:allerr} The dominating
 systematic uncertainties in the \GGG\, product for three scans (\%).
The correlated parts of the uncertainties are also presented.
\textcolor{black}{The inaccuracy of about 0.9\% due to possible interference 
phase correlation is not included.}
}}
\begin{center}
\begin{tabular}{llllll}
\emph{Source}
          & \multicolumn{1}{c}{Scan 1} &  \multicolumn{1}{c}{Scan 2} &
            \multicolumn{1}{c}{Scan 3} & 
           \multicolumn{1}{c}{Common$_{12}$} & 
            \multicolumn{1}{l}{\hspace*{-0.5em}Common$_{123}$} \\*[2pt]
\hline
 \textcolor{black}{Absolute} luminosity measurements 
                & \BL 1.6  &\BL  1.7 &\BL 1.2& \BL\BL1.6  &\BL 0.5         \\  \hline
 \PP decay simulation 
                & \BL 1.0 &\BL  1.0 &\BL 1.1& \BL\BL1.0  &\BL 1.0   \\  \hline
 Detector response \\ \hline 
 \hspace*{1em}Trigger efficiency 
                & \BL  0.2  &\BL  0.2 &\BL 0.2 & \BL\BL 0.2  &\BL 0.2 \\ \hline
 \hspace*{1em}Nuclear interaction  
                 & \BL 0.2 &\BL  0.2 &\BL 0.3 & \BL\BL 0.2  &\BL 0.2 \\  \hline
 \hspace*{1em}Cross talks in VD  
               & \BL 0.1  & \BL 0.17&\BL 0.1& \BL\BL0.1 &\BL 0.1    \\  \hline
 \hspace*{1em}Variation of cuts
                & \BL 0.5  &\BL  0.3 &\BL 0.6& \BL\BL0.3  &\BL 0.3   \\  \hline 
 Accelerator related effects \\ \hline 
 \hspace*{1em}Beam energy determination 
               & \BL 0.15 & \BL 0.18&\BL 0.6& \BL\BL0.15&\BL 0.15     \\  \hline
 \hspace*{1em}\textcolor{black}{Non-Gaussian energy distribution}
               & \BL 0.2 & \BL 0.2&\BL 0.2& \BL\BL0.2&\BL 0.2     \\  \hline
 \hspace*{1em}\textcolor{black}{Residual background}
               & \LT 0.1 & \LT 0.1&\LT 0.1& \BL\LT0.1&\LT 0.1     \\  \hline
 \textcolor{black}{Other uncertainties}
               & \BL 0.3 & \BL  0.3 &\BL0.3&\BL\BL0.3&\BL 0.3  \\  \hline\hline
\emph{Sum in quadrature}   
               & \AP 2.0  &\AP 2.1& \AP 1.9 &\BL\AP 2.0 &\AP 1.3 \\
\end{tabular}
\end{center}
\end{table*}

We repeated the procedure described above with the alternative set of event
selection criteria and obtained a slightly different uncertainty
estimate. To account for this we introduced an additional error of 0.3\%.

The contributions to the detection efficiency uncertainty
due to imperfect simulation of \PP decays are summarized in
Table~\ref{tab:mcerr}.

\subsection{Detector- and accelerator-related uncertainties in \GGG}
\label{err:detector}

The major sources of the systematic uncertainty in the \GGG\, product
are listed in Table~\ref{tab:allerr}.

The systematic error related to the efficiency of the track reconstruction
was considered in the previous section. The trigger
efficiency uncertainty is mainly due to
the uncertainty of the calorimeter thresholds in the secondary trigger.
The estimate of about $0.2\%$ was obtained varying the threshold
in the software event filter. \textcolor{black}{The excess of the
filter threshold over the hardware one in units of
the hardware threshold width was varied from 5 to 6.}

The trigger efficiency for all scans and the event selection efficiency for the
third one depend on the calorimeter response to hadrons. Their 
nuclear interaction
was simulated employing the GEISHA~\cite{Fesefeldt:1985yw} and
FLUKA~\cite{Fasso:2005zz} packages as they are implemented
in GEANT~3.21~\cite{GEANT:Tool}. Both of them satisfactorily reproduce
the pion signal in CsI, but in liquid krypton the performance of GEISHA
is much better, thus we determined the efficiencies using it and
estimated the systematic errors comparing the results obtained
with two packages.

The crosstalks in the vertex detector electronics introduced a
variation of the detection efficiency up to 1.5\% because
of the cut on the total number of VD hits in the on-line event selection
software (Section~\ref{sec:VEPP}). A
code simulating the crosstalks was developed and tuned
using either events of Bhabha scattering or cosmic rays.
The residual uncertainty thus determined is about 0.1$\div$0.2\%
depending on the VD voltage.

The effect of other possible sources of the detector-related
uncertainty was evaluated
by varying the event selection cuts.
The conditions on the number of tracks were \textcolor{black}{tightened
one by one} for all
three scans. For the last one a cut on the energy deposited
in the calorimeter was also \textcolor{black}{increased by
the most probable photon energy}.
The detection efficiency varied from 0.53 to 0.63 in the 2004
scans and from 0.70 to 0.79 in the 2006 scan. The maximum
variations of the \GGG\, result are presented in Table~\ref{tab:allerr}.

The systematic error in the \GGG\, value due to the beam energy determination
(Ref.~\cite{Blinov:2009})
and the data point formation (Section~\ref{sec:fit}) was studied
for each scan. The relative error does not exceed 0.2\%
except the third scan which includes one point with a significant 
accelerator instability that increases the error up to $0.6\%$.

\textcolor{black}{
The non-Gaussian effects in the total
collision energy distribution contribute
about 0.2\% to the \GGG\, uncertainty.
Changing the zero value of the $a$ parameter entering the preexponential
factor~\eqref{eq:PreExp} to the value measured with
the specific accelerator technique leads to the mass
shift of a few keV~\cite{Aulchenko:2003qq} and causes a negligible
bias in the \GGG\, value which is related to the area under the resonance
excitation curve. The quoted estimate was obtained by
releasing the $b$ parameter.}

\textcolor{black}{The bias in the resonance parameters
due to the admixture of the machine and cosmic background
to selected multihadron events was evaluated with the
controllable increase of this admixture. For each data point
the background event sample was formed containing
the events which passed some loose selection
criterea but were rejected by the multihadron ones. The number of multihadrons
events was increased by such fraction of the background sample size,
that the data fit yieled 2\% (0.4\%) growth of the continuum cross section
for the third (first and second) scan in
accordance with the backround estimates quoted in
Section~\ref{sec:trigger}.
The variation of the \PP mass value
did not exceed 1 keV, the change of the \GGG\, product was less
than 0.1\%.}

\subsection{\textcolor{black}{Other uncertainties}}
\label{err:fit}
In this subsection we discuss the uncertainties related to the
iterations used to obtain the total and electron width values
(Section~\ref{sec:theory}),
the fixation of the interference parameter $\lambda$ entering
the multihadron cross section~\eqref{BWrelativistic},
and the accuracy of the theoretic formulae employed for the
calculation of the cross section.
Using various initial values for the total  and 
electron width we have verified that the iteration procedure
converges \textcolor{black}{fast introducing a negligible systematic
uncertainty.}
To reduce the statistical error on the \PP mass,
the interference parameter $\lambda$
was fixed in the fit at the value of 0.13 corresponding to
Eq.~\eqref{eq:lamhtau}.
\textcolor{black}{
Releasing the $\lambda$ parameter in the fit shifts
the \GGG\, value by -0.23\%. This quantity can be used as an
estimate of the influence of quasi-two-body decays with
correlated interference phases mentioned in Section~\ref{Sec:Interference}.}
The accuracy of the resonance term in Eq.~\eqref{BWrelativistic}
is about 0.1\% (Section~\ref{sec:theory}), and another 0.1\% should be
added because of the accuracy
of radiative correction calculations in Ref.~\cite{KF}. 
\textcolor{black}{The quadratic sum of these three contributions
is about 0.3\%.}

\section{\textcolor{black}{Inaccuracy due to interference in hadronic
cross section}}
\label{err:inter}
\textcolor{black}{
The fits done with a floating interference parameter gave
$$ \lambda=0.21 \pm 0.07 \pm 0.05$$
The systematic uncertainty is 
mainly due to the beam energy determination and stability of the
cross section measurement.
This result does not contradict to the assumption of
the uncorrelated interference phases (Section~\ref{Sec:Interference})
but can still indicate
the presence of some phase correlations.}

\textcolor{black}{
In order to evaluate a
possible deviation of the $\Gh$ value from the sum
of hadronic partial widths $\Gamma_h$,
we performed a Monte Carlo
simulation according to Eq.~\eqref{eq:Gh} and \eqref{eq:lambdaSum}.
The set of decay modes and the corresponding branching fractions were
obtained using the event generator described in Section~\ref{sec:MC},
with the number of different decay modes exceeding one thousand.
Damping of $\sin\phi_m$ and $\cos\phi_m$
because of the averaging $\left<\,\right>_{\Theta}$ was ignored.}

\textcolor{black}{
The inaccuracy in $\Gamma_h$ was estimated in the Bayesian approach. 
Possible sets of interference phases $\phi_m$ are characterized 
by the central value $\phi$ and the band width $\Delta\phi$.
It was assumed that probabilities of all $\phi$ and $\Delta\phi$
values are equal. The result is not sensitive to assumptions
on the band shape.
}

\textcolor{black}{
The Monte Carlo procedure consists of two steps.
At the first step in each Monte Carlo sampling
a set of $\phi_m$ was generated
for random $\phi$ and $\Delta\phi$ and
the corresponding $\lambda$ value was calculated.
The values of $\phi$, $\Delta\phi$ and $\lambda$ were saved
for the second step.
The first-step distribution in $\lambda$ is symmetric relative to the most
probable value of 0.13 corresponding to Eq.~\eqref{eq:lambda}.
At the second step an acception-rejection (Von Neumann's) method 
was employed to reproduce
the $\lambda$ distribution matching the results of the measurement:
the Gaussian with the average value of 0.21 and the width of 0.086.
The second-step distribution in $(\Gh-\Gamma_h)/\Gamma_h$ is
peaked at zero. It is not Gaussian but 68\% percent of the accepted 
samplings
are contained in the interval of $\pm0.009$, thus we concluded
that the inaccuracy due to possible interference phase correlations
in $\Gamma_{ee}\times\B_h$ value is
about 0.9\%. Since the Bayesian approach was employed,
appearance of the new information on the interference
in the hadronic cross section
can change this estimate.
}

\textcolor{black}{
It should be noted that the inaccuracy estimated in this section
is not specific for our results on the \PP partial and total widths
but is shared by many results obtained in other experiments
using the \PP multihadron cross section. The most precise
of them is the result of CLEO on the $\PP\to J/\psi\,\pi^{+}\pi^{-}$
branching fraction~\cite{MENDEZ08}, its quoted accuracy is 2.2\%.}
\textcolor{black}{That concerns, in particular, the
works~\cite{BESelwid,Ablikim:2006zq}.}

\section{Averaging of scan results}
The systematic errors on the \GGG\, values for three scans
and the estimates of their correlations are presented in
Table~\ref{tab:allerr}. 
The correlation of errors is a difficult issue.
In non-obvious cases the most conservative approach was used assuming
that the correlated part corresponds to the minimal uncertainty in
scans for a given uncertainty source.

To obtain the resulting value of the \GGG\, product, the scans were 
treated as independent experiments. 
The individual \GGG\, values were weighted using
their statistical errors and uncorrelated parts of systematic errors. 
Such procedure takes into account the random behaviour of uncorrelated
systematic errors thus converting them to statistical. 
Correspondingly, the systematic errors of individual scans reduce 
to their common part. 
The formal weighting recipe for the 
parameter \GGG\, is given below:
\begin{eqnarray}\nonumber
\langle \Gamma_{ee}\!\times\!\B_h\rangle & = &
 \sum w_{i}\!\cdot\! \left(\Gamma_{ee}\!\times\!\B_h \right)_{i} \,,
\nonumber \\ 
 \sigma^2_{\mathrm{stat}}& =&\sum w^2_{i}\!\cdot\!\sigma^2_{\mathrm{stat},i}\,, 
\nonumber \\ 
\sigma^2_{\mathrm{syst}} & =&\!\sum w^2_{i}\!\cdot\!(\sigma^2_{\mathrm{syst,i}}\!-\!\sigma^2_{\mathrm{syst},0})
 +\sigma^2_{\mathrm{syst},0} \,,\nonumber \\
w_i& = &1/(\sigma^2_{\mathrm{stat},i}\!+\!\sigma^2_{\mathrm{syst},i}\!-\!\sigma^2_{\mathrm{syst},0})\,,
\label{eq:recipe}
\end{eqnarray}
where $\sigma_{\mathrm{syst},0}$ denotes a common part of systematic
uncertainties. The recipe preserves the total error of the result.
\section{Summary}
The parameters of \PP  have been measured using the data 
collected with the KEDR detector at the VEPP-4M \ee collider in 2004 and 2006.
Our final result for the \GGG\, product is

\begin{equation*}
    \GGG  =  2.233  \pm 0.015  \pm 0.037 \pm 0.020 \,\text{keV}.  
\end{equation*}
\textcolor{black}{The third error quoted is an estimate of the
\textcolor{black}{model dependence of the result}
due to assumptions on the interference effects
in the cross section of \textcolor{black}{the single-photon} $e^{+}e^{-}$
annihilation to hadrons explicitly
considered in this work. 
Implicitly, the
same assumptions were employed to obtain the charmonium
leptonic width and the absolute branching fractions
in many experiments.}
This quantity was measured in several experiments but only the result
of \mbox{MARK-I}~\cite{abrams}, an order of magnitude less precise than ours, 
was published in such a form.
Usually  the \GGG\, product is converted to the electron
width value using existing results on the branching fraction to hadrons
$\B_h$ or the leptonic branching fractions.

Using the world average values of the electron and hadron branching
fractions from PDG~\cite{PDG} we obtained 
the electron partial width and the total width of \PP:
\begin{equation*}  
 \begin{split}\
\Gamma_{ee}& = 2.282 \pm 0.015 \pm 0.038 \pm 0.021 \,\text{keV} \,, \\
\Gamma & = 296  \pm 2 \pm 8 \pm 3 \,\text{keV} \,. \\
 \end{split}
\end{equation*}
These results are consistent with and 
more than two times more precise than any  of the previous experiments.

\textcolor{black}{The result on the \PP mass obtained in this work
\begin{equation*}
    M = 3686.114 \pm 0.007 \pm 0.011 \:\,^{+0.002}_{-0.012} \:\:\text{MeV}.
\end{equation*}
The statistical uncertainty is significantly reduced compared to that
reached at VEPP-4M in 2002, the systematic one is approximately the same.
Since the systematic and model errors are 
correlated, the combined analysis of the 2002, 2004 and 2006 data 
has to be performed.
It will be described in a dedicated paper the result
of which should supersede the results presented
above and in Ref.~\cite{Aulchenko:2003qq}. The reduction of the model
dependence is expected.}

\section*{Acknowledgments}
We greatly appreciate the efforts of our VEPP-4M colleagues to provide 
good  operation of the accelerator complex and the staff of experimental
laboratories for the permanent support in preparing and performing
this experiment. The authors are grateful to 
\textcolor{black}{V.~P. Druzhinin, P. Wang and C.~Z. Yuan 
for useful discussions
and thank V.~S. Fadin for a verification of results presented in
Section~\ref{sec:param}.}

The  work is supported in part by the Ministry of Education and
Science of the Russian Federation and grants Sci.School 6943.2010.2,
RFBR 10-02-00695, 11-02-00112, 11-02-00558.

\end{document}